\documentclass[aps,twocolumn,prx,superscriptaddress,showpacs,tightenlines]{revtex4-1}
\usepackage{amsmath}
\usepackage{graphicx}
\usepackage{epsfig}
\usepackage{amsfonts}
\usepackage{color}
\usepackage{epstopdf}
\usepackage{hyperref}
\hypersetup{hypertex=true,
            colorlinks=true,
            linkcolor=blue,
            anchorcolor=blue,
            citecolor=blue}

\begin{document}

\newcommand{\nn}{\nonumber}
\newcommand{\ms}[1]{\mbox{\scriptsize #1}}
\newcommand{\msi}[1]{\mbox{\scriptsize\textit{#1}}}
\newcommand{\dg}{^\dagger}
\newcommand{\smallfrac}[2]{\mbox{$\frac{#1}{#2}$}}
\newcommand{\Tr}{\text{Tr}}
\newcommand{\ket}[1]{|#1\rangle}
\newcommand{\bra}[1]{\langle#1|}
\bibliographystyle{apsrev}
\newcommand{\pfpx}[2]{\frac{\partial #1}{\partial #2}}
\newcommand{\dfdx}[2]{\frac{d #1}{d #2}}
\newcommand{\half}{\smallfrac{1}{2}}
\newcommand{\s}{{\mathcal S}}
\newcommand{\jord}{\color{red}}
\newcommand{\kurt}{\color{blue}}

\title{ The driven-Markovian master equation based on the Lewis-Riesenfeld invariants theory}

\author{S. L. Wu }
\email{slwu@dlnu.edu.cn}
\affiliation{School of Physics and Materials Engineering,
Dalian Nationalities University, Dalian 116600 China}

\author{X. L. Huang}
\affiliation{School of Physics and Electronic Technology,
Liaoning Normal University, Dalian 116029, China}

\author{X. X. Yi }
\email{yixx@nenu.edu.cn}
\affiliation{Center for Quantum Sciences and School of Physics,
Northeast Normal University, Changchun 130024, China}

\date{\today}

\begin{abstract}
We derive a  Markovian master equation for driven open quantum systems based on the Lewis-Riesenfeld
invariants theory, which is available for arbitrary driving protocols.The role of the Lewis-Riesenfeld
invariants is to help us bypass the time-ordering obstacle in expanding the propagator of the free
dynamics, such that the Lindblad operators in our driven-Markovian master equation can be determined
easily.  We also illustrate that, for the driven open quantum systems,  the spontaneous emission and the thermal
excitation induce the transitions between eigenstates of the Lewis-Riesenfeld invariant, but not
the system Hamiltonian's. As an example, we present the driven-Markovian master equation
for a driven two-level system coupled to a heat reservoir. By comparing to the exactly solvable models,
the availability of the driven-Markovian master equation is verified.  Meanwhile, the adiabatic limit and inertial limit
of the driven-Markovian master equation are also discussed, which result in the same Markovian master equations
as those presented before in the corresponding limits.
\end{abstract}

\maketitle

\section{Introduction}

Any physical system in nature, no matter classical or quantum, couples to its
surroundings exchanging energy and matter to make it as an open system.
The theory of open quantum systems aims at providing a concise manner to
describe the dynamics of the primary system\cite{Breuer2007}.  For open quantum
systems satisfying the Born-Markov approximation\cite{Davies1974},  the
Gorini-Kossakowski- Lindblad-Sudarshan (GKLS) Markovian master equation
gives a general completely positive and trace preserving
map of the reduced dynamics \cite{Gorini1976,Lindblad1976}. In the original
derivation, it is assumed that the system Hamiltonian is time-independent. The coupling
to the environment induces transitions between the static eigenstates of the system
Hamiltonian. For the open systems with time-dependent external drives, the GKLS
master equations have been derived in and beyond the adiabatic limit, which leads to the
adiabatic \cite{Davies1978,Albash2012,Childs2001,Kamleitner2013,Sarandy2004} and
non-adiabatic \cite{Yamaguchi2017,Dann2018,Potts2021} Markovian master equation.

For the driven open quantum systems without memory of the driving protocol, a
non-adiabatic Markovian master equation (NAME) has been derived \cite{Dann2018}. The Lindblad
operators in this non-adiabatic master equation are eigenoperators of the propagator
of the free dynamics which associates with the system Hamiltonian. Generally speaking, these
eigenoperators can be determined by representing the dynamics in the operator space which
is also known as the Liouvillian space \cite{Albert2016,Scopa2019}. However, it is difficult to
give these eigenoperators explicitly, and these eigenoperators  absent clear physical meanings .
For second best thing, a method based on the inertial theorem is proposed\cite{Dann2020}.
The inertial theorem relies on a priori decomposition of the Liouvillian
superoperator into a rapid-changing scalar parameter and a  slow-changing superoperator,
which is equivalent to additional restrictions on the driving protocol \cite{Dann2021}.
In this way, the Lindblad operators can be obtained explicitly, if the inertial limit is
reached\cite{Dann2020,Dann2021,Dann2019}. {At the same time, effective numerical
methods to simulate the driven open quantum system dynamics are also proposed
\cite{Hwang2012, Stockburger2016}, but may provide less structural insight into
the dynamics.}


For the open quantum systems with a static Hamiltonian, the population transitions
caused by decoherence occur between the eigenstates of the static Hamiltonian \cite{Breuer2007}.
Hence, even if the Markovian equation can not be given explicitly, we may formulate
a phenomenological master equation due to the clearly physical meanings of the transitions
\cite{Ozaki2021,Taran2019}. However, it is totally different for the driven open quantum
systems with a non-adiabatic driving protocol, since there is no such a physical meaning for the
decoherence-induced transitions. Thus, it is natural  to ask  if there is a simple method to formulate
a Markovian  master equation for a driven open quantum system with arbitrary driving protocols
as we used in formulating the Markovian master equation with the static Hamiltonian?

In this paper, we derive a driven-Markovian master equation (DMME) for arbitrary driving
protocols by using the Lewis-Riesenfeld invariants (LRIs) \cite{Lewis1968}, which is easy to
formulate and has a clear meaning of the decoherence-induced transitions. Since the solution
of the Shr\"{o}dinger equation with
a time-dependent Hamiltonian can be expressed as a superposition of the eigenstates of the
Lewis-Riesenfeld invariant with constant amplitudes \cite{Lewis1969}, the unitary operator
corresponding to the free propagator can be written down explicitly. On the other hand, if the
the timescale for the driving protocol, also known as "the non-adiabatic timescale",
approaches to or is larger than the reservoir correlation time, the memory effect of the
driving protocol can not be neglected. By using the Fourier transformation and
its inverse transformation, we collect the frequency contribution caused by the driving
memory effect into the one-side Fourier transformation of the reservoir correlation
function. Thus, the driving memory effect can be included in the DMME
with a concise manner. As we shall see, the transitions induced by the coupling to the environment
only occur between the invariant's eigenstates. Therefore, it is quiet straightforward to formulate
a DMME for a general driven open system, if its LRIs are known.

This paper is organized as follows. In Sec. \ref{sec:gf}, we present the general formula of
the DMME based on the Lewis-Riesenfeld invariants theory. The memory of the driven protocol
is encoded in the decoherence rates and the Lamb shifts of the DMME.  Then,
we apply this DMME to the two-level system with  time-dependent driving fields in Sec.
\ref{sec:2qubits}, and the corresponding adiabatic and inertial limits are discussed.
We also derived  exact dynamics for the driven two-level system, which will
help to illustrate validity and availability of the DMME. Finally, the conclusions are given in
Sec. \ref{sec:Conc}.

\section{General Formalism} \label{sec:gf}

In this section, we apply the LRIs theory to present the DMME with explicit mathematical
and physical meaning. Consider the dynamics of the total system which is governed by
the Hamiltonian
\[H(t)=H_{\text{s}}(t)+H_{\text{B}}+H_{\text{I}}.\]
$H_{\text{s}}(t)$ stands for the system Hamiltonian; the reservoir is represented by the
Hamiltonian
\[
H_{\text{B}}=\sum_{k}\omega_{k}b_{k}^{\dagger}b_{k},
\]
{where $b_k$ and $\omega_k$ are, respectively, the annihilation operator
and the eigenfrequency of the $k$-th mode of the reservoir.}
In the following, the natural units $\hbar=c=1$ are used throughout.
We assume that the system-reservoir interaction Hamiltonian is given by
\[
H_{\text{I}}=\sum_{k}\textsl{g}_{k}A_{k}\otimes B_{k}.
\]
$A_{k}$ and $B_{k}$ are the Hermitian operators of the system and reservoir, respectively.
$g_{k}$ stands for the coupling strength.

The von Neumann equation for the density operator of the total system
in the interaction picture reads
\[
\partial_{t}\tilde{\rho}(t)=-i\left[\tilde{H_{\text{I}}}(t),\tilde{\rho}(t)\right],
\]
 where $\tilde{\rho}(t)$ denotes the density operator of the total
system in the interaction picture, and a similar notations is applied
for the other system and reservoir operators. By assuming the weak system-reservoir
coupling (the Born approximation), we obtain the Born equation for
the system density operator $\tilde{\rho}_{\text{s}}(t)$,
\begin{eqnarray*}
\partial_{t}&&\tilde{\rho}_{\text{s}}(t)=\nonumber\\
&&-\int_{0}^{t}d\tau\text{Tr}_{\text{B}}\left\{ \left[\tilde{H_{\text{I}}}(t),\left[\tilde{H_{\text{I}}}(t-\tau),
\tilde{\rho}_{\text{s}}(t-\tau)\otimes\tilde{\rho}_{\text{B}}\right]\right]\right\} .
\end{eqnarray*}
 Here, we have assumed that $\text{Tr}_{\text{B}}\{ \tilde{H_{\text{I}}}(t)\tilde{\rho}(0)\} =0$,
and the initial state of the total system can be written as $\tilde{\rho}(0)=
\tilde{\rho}_{\text{s}}(0)\otimes\tilde{\rho}_{\text{B}}$. The Born approximation
assumes that the coupling between the system and the reservoir is weak,
such that the influence of the system on the reservoir is small.
If the system evolution time is much larger than the reservoir correlation
time $\tau_{\text{B}}$, we can replace $\tilde{\rho}_{\text{s}}(t-\tau)$
by $\tilde{\rho}_{\text{s}}(t)$ and the integral limits can be extended
to $\infty$, which is known as the Markovian approximation. In such a
case, the dynamics governed by the following Redfield master
equation within the Born-Markovian approximation \cite{Redfield1965},
\begin{eqnarray}
\partial_{t}&&\tilde{\rho}_{\text{s}}(t)=\nonumber\\
&&-\int_{0}^{\infty}d\tau\text{Tr}_{\text{B}}\left\{ \left[\tilde{H_{\text{I}}}(t),\left[\tilde{H_{\text{I}}}(t-\tau),
\tilde{\rho}_{\text{s}}(t)\otimes\tilde{\rho}_{\text{B}}\right]\right]\right\} .\label{eq:rfmequ}
\end{eqnarray}
For an operator $A$ of the system, the corresponding operator in the
interaction picture can be connected by an unitary transformation, i.e.,
\begin{eqnarray}
\tilde{A}_{k}(t)=\hat{\mathcal{U}}_{\text{s}}(t)A_{k}=U_{\text{s}}^{\dagger}(t)A_{k}U_{\text{s}}(t).\label{at}
\end{eqnarray}
$U_{\text{s}}(t)$ describing the free propagator of the system satisfies
a Schr\"{o}dinger equation for the time-dependent system Hamiltonian
\begin{equation}
i\partial_{t}U_{\text{s}}(t)=H_{\text{s}}(t)U_{\text{s}}(t),\:U_{\text{s}}(0)=I,\label{eq:ueqn}
\end{equation}
which results in $U_{\text{s}}(t)=\mathcal{T}\exp\left(-i\int_{0}^{t}\text{d}\tau\,H_{\text{s}}(\tau)\right)$
with the time-ordering operator $\mathcal{T}$.

To reduce Eq.(\ref{eq:rfmequ}), a set of eigenoperator of the superoperator
$\mathcal{\hat{U}}_{\text{s}}(t)$ are needed, where the eigenoperators satisfy
$\tilde{F}_{j}(t)=\hat{\mathcal{U}}_{\text{s}}(t)\tilde{F}_{j}(0)=\lambda_{j}(t)\tilde{F}_{j}(0)$
\cite{Dann2018}. However, it is difficult to solve the eigenequation
of the superoperator $\mathcal{\hat{U}}_{\text{s}}(t)$ directly.
To overcome this difficulty, the inertial theorem has been used to
obtain an approximative solution of $\hat{\mathcal{U}}_{\text{s}}(t)$
\cite{Dann2021,Dann2020}. However, the solution is accurate only
under that the inertial parameter is small, which requires a slow
acceleration of the drive.

In fact, the free propagator of the system can be obtained directly,
if the Lewis-Riesenfeld invariants for the system Hamiltonian $H_{\text{s}}(t)$
are known. A LRI $I_{\text{s}}(t)$ for the systems with the Hamiltonian
$H_{\text{s}}(t)$ is a Hermitian operator which obeys  an equation in the
Schr\"{o}dinger picture \cite{Lewis1969}
\begin{equation}
i\partial_{t}I_{\text{s}}(t)-\left[H_{\text{s}}(t),I_{\text{s}}(t)\right]=0.\label{eq:dIs}
\end{equation}
For the closed dynamics
of the systems, a general solution of the Schr\"{o}dinger equation can
be written as
\begin{equation}
|\Psi(t)\rangle=\sum_{n=1}^{N}c_{n}\text{e}^{i\alpha_{n}(t)}|\psi_{n}(t)\rangle.\label{eq:psit}
\end{equation}
Here, $|\psi_{n}(t)\rangle$ is the $n$-th eigenstate of the LRI
with a real constant eigenvalue $\lambda_{n}$, i.e.,
$I_{\text{s}}(t)|\psi_{n}(t)\rangle=\lambda_{n}|\psi_{n}(t)\rangle$,
$\{c_{n}\}$ are time-independent amplitudes, and the Lewis-Riesenfeld
phases are defined as \cite{Lewis1969}
\begin{equation}
\alpha_{n}(t)=\int_{0}^{t}\langle\psi_{n}(\tau)|\left(i\partial_{\tau}
-H_{\text{s}}(\tau)\right)|\psi_{n}(\tau)\rangle\,\text{d}\tau.\label{eq:lrp}
\end{equation}
Therefore, the solution of Eq.(\ref{eq:ueqn}) can be expressed by
means of the eigenstates of the LRIs,
\begin{equation}
U_{\text{s}}(t)=\sum_{n}\text{e}^{i\alpha_{n}(t)}|\psi_{n}(t)\rangle\langle\psi_{n}(0)|.\label{eq:us}
\end{equation}

{The LRIs theory was designed to investigate the time evolution
of dynamical systems with an explicitly time-dependent Hamiltonian
\cite{Lewis1969}.  The invariants comply with the following properties:
(i) The expectation values of the LRIs are constant. (ii)
The eigenvalues of a LRI are constant, while the eigenstates
are time dependent. (iii) Any time-dependent Hermitian operator which
satisfies Eq. (\ref{eq:dIs}) is a LRI for closed quantum
systems. Each LRI corresponds to a symmetry of the closed
quantum system. Thereafter, the LRIs are successfully
applied to investigate time-dependent problems in quantum mechanics
\cite{Pedrosa2009} such as the Berry phase \cite{Sarandy2007}, the connection
between quantum theory  and classical theory \cite{Schuch2006}, and the
quantum control \cite{Chen2011}.
At the same time, the method to construct the LRIs for various
quantum systems has been explored, for instance, the harmonic oscillator system
\cite{Lewis1969}, the few-level systems \cite{Monteoliva1994,Nakahara2012},
the pseudo-Hermitian system\cite{Simeonov2016}, and the open fermionic
systems\cite{Kim2000}. Also a general method for constructing LRIs has been
proposed\cite{Ponte2018}.}

By means of the explicit formula of the free evolution operator $U_{\text{s}}(t)$ (Eq.(\ref{eq:us})),
the system operator Eq.(\ref{at})  in the interaction picture  can be rewritten as
\begin{eqnarray}
\tilde{A}_{k}(t) & = & U_{\text{s}}^{\dagger}(t)A_{k}U_{\text{s}}(t)\nonumber \\
 & = & \sum_{n,m}\text{e}^{i\theta_{mn}^{k}(t)}\xi_{mn}^{k}(t)\tilde{F}_{mn}\label{eq:Fj}
\end{eqnarray}
with
\begin{eqnarray}
\theta_{mn}^{k}(t)=\alpha_{n}(t)-\alpha_{m}(t)+\text{Arg}
\left(\langle\psi_{m}(t)|A_{k}|\psi_{n}(t)\rangle\right)\label{eq:if}
\end{eqnarray}
and $\xi_{mn}^{k}(t)=|\langle\psi_{m}(t)|A_{k}|\psi_{n}(t)\rangle|$ which satisfy
$\theta_{mn}^{k}(t)\in\mathbb{R}$ and $\xi_{mn}^{k}(t)>0.$
The time-independent operators $\tilde{F}_{mn}=|\psi_{m}(0)\rangle\langle\psi_{n}(0)|$
denotes one of Lindblad operators in the interaction picture. Since $\tilde{A}_{k}(t)$ are Hermitian
operators, it yields
\begin{equation}
\tilde{A}_{k}(t)=\sum_{n',m'}\text{e}^{-i\theta_{m'n'}^{k}(t)}\xi_{m'n'}^{k}(t)\tilde{F}_{m'n'}^{\dagger}.\label{eq:Fjd}
\end{equation}
Any $\tilde F_{m'n'}^{\dagger}$ contains in the operator set $\left\{ \tilde F_{mn}\right\} $
which  can be used to expand the corresponding Liouvillian space\cite{Petrosky1997}.
Substituting Eqs.(\ref{eq:Fj}) and (\ref{eq:Fjd}) into Eq.(\ref{eq:rfmequ}),
we can express the Markovian master equation as
\begin{eqnarray}
\partial_{t}\tilde{\rho}_{\text{s}}  (t)&= & \sum_{m,m',n,,n'}\Gamma_{mn,m'n'}(t)
\left(\tilde{F}_{mn}\tilde{\rho}_{\text{s}}(t)\tilde{F}_{m'n'}^{\dagger}\right.\nonumber\\
&&\left.-\tilde{F}_{m'n'}^{\dagger}\tilde{F}_{mn}\tilde{\rho}_{\text{s}}(t)\right)+\text{H.c.}
\label{meq1}
\end{eqnarray}
with
\begin{eqnarray}
\Gamma_{mn,m'n'}&&(t)=\sum_{k,k'}\textsl{g}_{k}
\textsl{g}_{k'}\nonumber\\\times&& \int_{0}^{\infty}\text{d}s\xi_{m'n'}^{k'}(t)\xi_{mn}^{k}(t-s)
\text{e}^{i\left(\theta_{mn}^{k}(t-s)-\theta_{m'n'}^{k'}(t)\right)}\nonumber\\
 \times && \text{Tr}_{\text{B}} \left\{\tilde{B}_{k'}(t)\tilde{B}_{k}(t-s)\rho_{B}\right\},
 \label{gam1}
\end{eqnarray}
where H.c. denotes the Hermitian conjugated expression and $\tilde{B}_{k'}(t)$
is the bath operator in the interaction picture.

As shown in Eq.(\ref{meq1}), there is memory effect of the driving
protocol, which contains in $\xi_{mn}^{k}(t-s)$ and $\theta_{mn}^{k}(t-s)$.
At first, {by means of the Taylor expansion, the phase $\theta_{mn}^{k}(t-s)$
can be written as
\begin{eqnarray}
\theta_{mn}^{k}(t-s) & = & \theta_{mn}^{k}(t)+\partial_{s}\theta_{mn}^{k}(t-s)|_{s=0}s\nonumber \\
 &  & +\sum_{l=2}^{\infty}\frac{1}{l!}\partial_{s}^{l}\theta_{mn}^{k}(t-s)s^{l}\nonumber\\
 & \equiv & \theta_{mn}^{k}(t)+\alpha_{mn}^{k}(t)s+\varTheta_{mn}^{k}(t,t-s),\label{eq:thetas-1}
\end{eqnarray}
where
\begin{eqnarray}
\varTheta_{mn}^{k}(t,t-s)&=&\theta_{mn}^{k}(t-s)-\theta_{mn}^{k}(t)-\alpha_{mn}^{k}(t)s\nonumber\\
&=& \int_{t-s}^t (\alpha_{mn}^{k}(\tau)-\alpha_{mn}^{k}(t))d\tau
\end{eqnarray}
is a function of $t$ and $t-s$ with $\alpha_{mn}^{k}(t)=-\partial_{t}\theta_{mn}^{k}(t)$. }With the
consideration of $\text{e}^{i\varTheta_{mn}^{k}}=\cos\varTheta_{mn}^{k}+i\sin\varTheta_{mn}^{k}$,
Eq.(\ref{gam1}) becomes
\begin{eqnarray}
\Gamma&&_{mn,m'n'}(t)\nonumber\\ & &= \sum_{k,k'}\textsl{g}_{k}\textsl{g}_{k'}\xi_{m'n'}^{k'}(t)
\text{e}^{i\left(\theta_{mn}^{k}(t)-\theta_{m'n'}^{k'}(t)\right)}\nonumber\\
&&\times\int_{0}^{\infty}\text{d}s\left(\Xi_{mn}^{\text{c},k}(t,t-s)+i\Xi_{mn}^{\text{s},k}(t,t-s)\right)\nonumber \\
 && \times  \text{Tr}_{\text{B}}\left\{\tilde{B}_{k'}(t)\tilde{B}_{k}(t-s)\rho_{B}\right\}
 \text{e}^{i\alpha_{mn}^{k}(t)s},\label{eq:gammamn}
\end{eqnarray}
where $\Xi_{mn}^{\text{c},k}(t,t-s)=\xi_{mn}^{k}(t-s)\cos\varTheta_{mn}^{k}(t,t-s)$
and $\Xi_{mn}^{\text{s},k}(t,t-s)=\xi_{mn}^{k}(t-s)\sin\varTheta_{mn}^{k}(t,t-s)$.
For $\Xi_{mn}^{\text{c(s)},k}(t,t-s)$, we take the Fourier expansion
with respect to $t-s$,
\begin{equation}
\Xi_{mn}^{\text{c(s)},k}(t,t-s)=\frac{1}{\sqrt{2\pi}}\int_{-\infty}^{+\infty}d\omega_{\xi}
\bar{\Xi}_{mn}^{\text{c(s)},k}(t,\omega_{\xi})\text{e}^{i\omega_{\xi}(t-s)}\label{eq:xis}
\end{equation}
with $\bar{\Xi}_{mn}^{\text{c(s)},k}(t,\omega_{\xi})=\frac{1}{\sqrt{2\pi}}
\int_{-\infty}^{+\infty}\Xi_{mn}^{c(s),k}(t,\tau)\text{e}^{-i\omega_{\xi}\tau}d\tau$.
By substituting Eq. (\ref{eq:xis}) into Eq.(\ref{eq:gammamn}), it
yields
\begin{eqnarray*}
\Gamma&&_{mn,m'n'}(t)\nonumber\\ & &= \frac{1}{\sqrt{2\pi}}\sum_{k,k'}\textsl{g}_{k}\textsl{g}_{k'}
\xi_{m'n'}^{k'}(t)\text{e}^{i\left(\theta_{mn}^{k}(t)-\theta_{m'n'}^{k'}(t)\right)}\\
 &&  \times \int_{-\infty}^{+\infty}d\omega_{\xi}\left(\bar{\Xi}_{mn}^{\text{c},k}(t,\omega_{\xi})
 +i\bar{\Xi}_{mn}^{\text{s},k}(t,\omega_{\xi})\right)\text{e}^{i\omega_{\xi}t}\\
 && \times  \bar{\varLambda}_{kk'}(\alpha_{mn}^{k}-\omega_{\xi}),
\end{eqnarray*}
where $\bar{\varLambda}_{kk'}$ is the one-sided Fourier transform
of the instantaneous reservoir correlation function
\begin{eqnarray}
\bar{\varLambda}_{kk'}(\alpha) & = & \int_{0}^{\infty}\text{d}s
\text{e}^{i\alpha s}\text{Tr}_{\text{B}}\left\{\tilde{B}_{k'}(t)\tilde{B}_{k}(t-s)\rho_{B}\right\}.\label{eq:gammak}
\end{eqnarray}
with $\alpha=\alpha_{mn}^{k}-\omega_{\xi}$. It is convenient to decompose $\bar{\varLambda}_{kk'}$ into a real
and imaginary part, i.e.,
\[
\bar{\varLambda}_{kk'}(\alpha)=\bar{\varLambda}_{kk'}^{\text{R}}(\alpha)+i\,\bar{\varLambda}_{kk'}^{\text{I}}(\alpha),
\]
 where $\bar\Lambda_{kk'}^{\text{I}}(\alpha)=-\frac{i}{2}\left(\bar\Lambda_{kk'}(\alpha)-\bar\Lambda_{kk'}^{*}(\alpha)\right)$
is a Hermitian matrix and $\bar\Lambda_{kk'}^{\text{R}}(\alpha)$ can
be written as
\[
\bar{\varLambda}_{kk'}^{\text{R}}(\alpha)=\frac{1}{2}\int_{-\infty}^{\infty}\text{d}s
\text{e}^{i\alpha s}\text{Tr}_{\text{B}}\left\{\tilde{B}_{k}(s)\tilde{B}_{k'}(0)\rho_{B}\right\}.
\]
 We divide $\Gamma_{mn,m'n'}$ into the real and imaginary
parts, i.e.,
\begin{widetext}
\begin{eqnarray}
\Gamma_{mn,m'n'} (t)& = & \frac{1}{\sqrt{2\pi}}\sum_{k,k'}\textsl{g}_{k}\textsl{g}_{k'}\xi_{m'n'}^{k'}(t)
\text{e}^{i\left(\theta_{mn}^{k}(t)-\theta_{m'n'}^{k'}(t)\right)}\int_{-\infty}^{+\infty}d\omega_{\xi}\text{e}^{i\omega_{\xi}t}\nonumber\\
 & \times & \left(\left(\bar{\Xi}_{mn}^{\text{c},k}(t,\omega_{\xi})\bar{\varLambda}_{kk'}^{\text{R}}(\alpha_{mn}^{k}
 -\omega_{\xi})-\bar{\Xi}_{mn}^{\text{s},k}(t,\omega_{\xi})\bar{\varLambda}_{kk'}^{\text{I}}(\alpha_{mn}^{k}-\omega_{\xi})\right)\right.\nonumber\\
 & + & \left.i\!\left(\bar{\Xi}_{mn}^{\text{c},k}(t,\omega_{\xi})\bar{\varLambda}_{kk'}^{\text{I}}(\alpha_{mn}^{k}
 -\omega_{\xi})+\bar{\Xi}_{mn}^{\text{s},k}(t,\omega_{\xi})\bar{\varLambda}_{kk'}^{\text{R}}(\alpha_{mn}^{k}-\omega_{\xi})\right)\right).\label{eq:gamf}
\end{eqnarray}
 According to the convolution theorem of the Fourier transformation,
we can transform the integral in $\Gamma_{mn,m'n'}$ with respect to
$\omega_{\xi}$ into a convolution of the time-domain integral, which
leads to
\begin{eqnarray*}
\Gamma_{mn,m'n'} (t)& = & \frac{1}{2\pi}\sum_{k,k'}\textsl{g}_{k}\textsl{g}_{k'}\xi_{m'n'}^{k'}(t)
\text{e}^{i\left(\theta_{mn}^{k}(t)-\theta_{m'n'}^{k'}(t)\right)}\int_{-\infty}^{+\infty}ds'\\
 & \times & \left(\left(\Xi_{mn}^{\text{c},k}(t,t-s')\Lambda_{kk'}^{\text{R}}(\alpha_{mn}^{k},s')
 -\Xi_{mn}^{\text{s},k}(t,t-s')\Lambda_{kk'}^{\text{I}}(\alpha_{mn}^{k},s')\right)\right.\\
 & + & \left.i\!\left(\Xi_{mn}^{\text{c},k}(t,t-s')\Lambda_{kk'}^{\text{I}}(\alpha_{mn}^{k},s')
 +\Xi_{mn}^{\text{s},k}(t,t-s')\Lambda_{kk'}^{\text{R}}(\alpha_{mn}^{k},s')\right)\right)
\end{eqnarray*}
\end{widetext}
with $\Lambda_{kk'}^{\text{R(I)}}(\alpha_{mn}^{k},s')=\frac{1}{\sqrt{2\pi}}\int_{-\infty}^{+\infty}
d\omega_{\xi}\text{e}^{i\omega_{\xi}s'}\bar{\varLambda}_{kk'}^{\text{R(I)}}(\alpha_{mn}^{k}-\omega_{\xi})$.

The system evolution time $\tau_{\text{s}}$ is the typical timescale
of the intrinsic evolution of the system, which is defined by a typical
value for the inverse of the instantaneous frequency differences involved,
i.e. $\tau_{\text{s}}\propto|\alpha_{mn}^{k}(t)-\alpha_{m'n'}^{k'}(t)|^{-1}$
for $\alpha_{mn}^{k}(t)\neq\alpha_{m'n'}^{k'}(t)$.
{If $\tau_{\text{s}}$ is small compared to the relaxation time $\tau_{\text{R}}$,}
the non-secular terms in the DMME with $\alpha_{mn}^{k}(t)\neq\alpha_{m'n'}^{k'}(t)$
may be neglected, which is known as the secular approximation. Thus
we have the DMME within the secular approximation in the interaction
picture,
\begin{eqnarray}
\partial_{t}\tilde{\rho}_{\text{s}}&&(t)  =  -i\left[\tilde{H}_{\text{LS}}(t),
\tilde{\rho}_{\text{s}}(t)\right]+\sum_{m,m',n,,n'}\Gamma_{mn,m'n'}^{\text{R}}(t)\nonumber\\
 && \times  \left(\tilde{F}_{mn}\tilde{\rho}_{\text{s}}(t)\tilde{F}_{m'n'}^{\dagger}-\frac{1}{2}
 \left\{ \tilde{F}_{m'n'}^{\dagger}\tilde{F}_{mn},\tilde{\rho}_{\text{s}}(t)\right\} \right), \label{eq:nonM}
\end{eqnarray}
 in which $\tilde{H}_{\text{LS}}(t)=\sum_{m,n,m',n'}\Gamma_{mn,m'n'}^{\text{I}}(t)\tilde{F}_{m'n'}^{\dagger}\tilde{F}_{mn}$
is the Lamb shift Hamiltonian and $\Gamma_{mn,m'n'}^{\text{R(I)}}(t)$ denotes
the real (imaginary) part of $\Gamma_{mn,m'n'}(t)$,
\begin{eqnarray}
&\Gamma&_{mn,m'n'}^{\text{R}}(t)  =  \frac{1}{\pi}\sum_{k}\textsl{g}_{k}^{2}
\xi_{m'n'}^{k}(t)\int_{-\infty}^{+\infty}ds' \label{eq:gamr}\\
 && \times  \left(\Xi_{mn}^{\text{c},k}(t-s')\Lambda_{kk}^{\text{R}}(\alpha_{mn}^{k},s')-
 \Xi_{mn}^{\text{s},k}(t-s')\Lambda_{kk}^{\text{I}}(\alpha_{mn}^{k},s')\right),\nonumber
\end{eqnarray}
\begin{eqnarray}
&\Gamma&_{mn,m'n'}^{\text{I}} (t)  =  \frac{1}{2\pi}\sum_{k}\textsl{g}_{k}^{2}
\xi_{m'n'}^{k}(t)\int_{-\infty}^{+\infty}ds'\label{eq:gami}\\
 && \times  \left(\Xi_{mn}^{\text{c},k}(t-s')\Lambda_{kk}^{\text{I}}(\alpha_{mn}^{k},s')
 +\Xi_{mn}^{\text{s},k}(t-s')\Lambda_{kk}^{\text{R}}(\alpha_{mn}^{k},s')\right).\nonumber
\end{eqnarray}

In fact, although we admit non-adiabatic change of the driving protocol,
the DMME presented in Eq.(\ref{eq:nonM}) describes a Markovian dynamics
if $\Gamma_{mn,m'n'}^{\text{R}}(t)\geq0$ for $\forall t$. The memory
effects of the driving protocol is explicitly encoded into a convolution
with the reservoir correlation function. { In order to connect to the
previous results\cite{Dann2018,Dann2019}, we may assume
that the change of the phase $\theta_{mn}^{k}(t)$
is slow comparing to the reservoir correlation decay rate.} Thus, there is
a typical timescale $\tau_{\text{d}}$, called ``the non-adiabatic
phase timescale'', defined as\cite{Dann2018}
\[
\tau_{\text{d}}\equiv\text{Min}_{m,n,k,t}\left\{ \frac{\partial_{t}\theta_{mn}^{k}(t)}
{\partial_{t}^{2}\theta_{mn}^{k}(t)}\right\} ,
\]
 which is related to the change of the phase in the driving protocol.
Thus, the assumption of the slow changing phase is equivalent to require
that the reservoir correlation time $\tau_{\text{B}}$ has to be much
smaller than the non-adiabatic phase timescale $\tau_{\text{d}}$,
i.e., $\tau_{\text{B}}\ll\tau_{\text{d}}$. For $s\in\left[0,\tau_{\text{B}}\right]$
and $s\ll t$, the terms up to second order in Eq.(\ref{eq:thetas-1})
can be ignored, i.e., $\varTheta_{mn}^{k}=0,$ so that $\Xi_{mn}^{\text{c},k}(t,t-s)=\xi_{mn}^{k}(t-s)$
and $\Xi_{mn}^{\text{s},k}(t,t-s)=0$. Therefore, the real (imaginary)
part of $\Gamma_{mn,m'n'}(t)$ becomes
\begin{eqnarray*}
\Gamma_{mn,m'n'}^{\text{R}}(t) & = & \frac{1}{\pi}\sum_{k}\textsl{g}_{k}^{2}\xi_{m'n'}^{k}(t)
\int_{-\infty}^{+\infty}d\omega_{\xi}\bar{\xi}_{mn}^{k}(\omega_{\xi})\nonumber\\
&&\times\bar{\varLambda}_{kk}^{\text{R}}(\alpha_{mn}^{k}-\omega_{\xi})\text{e}^{i\omega_{\xi}t}.\\
\Gamma_{mn,m'n'}^{\text{I}}(t) & = & \frac{1}{2\pi}\sum_{k}\textsl{g}_{k}^{2}\xi_{m'n'}^{k}(t)
\int_{-\infty}^{+\infty}d\omega_{\xi}\bar{\xi}_{mn}^{k}(\omega_{\xi})\nonumber\\
&&\bar{\varLambda}_{kk}^{\text{I}}(\alpha_{mn}^{k}-\omega_{\xi})\text{e}^{i\omega_{\xi}t}.
\end{eqnarray*}
 If the change of $\xi_{mn}^{k}(t)$ is much smaller than the instantaneous
frequency, i.e., $\omega_{\xi}\ll\alpha_{mn}^{k}$, we immediately
obtain
\begin{eqnarray}
\Gamma_{mn,m'n'}^{\text{R}}(t) & = & 2\sum_{k}\textsl{g}_{k}^{2}\xi_{m'n'}^{k}(t)\xi_{mn}^{k}(t)
\bar{\varLambda}_{kk}^{\text{R}}(\alpha_{mn}^{k}),\nonumber\\
\Gamma_{mn,m'n'}^{\text{I}}(t) & = & \sum_{k}\textsl{g}_{k}^{2}\xi_{m'n'}^{k}(t)\xi_{mn}^{k}(t)
\bar{\varLambda}_{kk}^{\text{I}}(\alpha_{mn}^{k}),\label{eq:gadi}
\end{eqnarray}
 which lead to the non-adiabatic Markovian master equation given in Ref.\cite{Dann2018}.

The DMME presented in Eq.(\ref{eq:nonM}) does not contain any approximation
on the driving protocol. It is interesting that both the real and
imaginary parts of the one-sided Fourier transform of the instantaneous
reservoir correlation function $\bar{\varLambda}_{kk}(\alpha)$ are
involved in both the Lamb shift and the decoherence (see Eqs. (\ref{eq:gamr})
and (\ref{eq:gami})). For the Markovian master equation with the
static Hamiltonian and the time-dependent Hamiltonian satisfying $\tau_{\text{B}}\ll\tau_{\text{d}}$,
$\bar{\varLambda}_{kk}^{\text{R}}(\alpha)$ only contributes to the
decoherence process, while $\bar{\varLambda}_{kk}^{\text{I}}(\alpha)$
just appears in the Lamb shift. As a result, the positive decoherence rates
may not be ensured and additional energy level shifts can be observed
in the driven open quantum systems for the timescale $\tau_{\text{B}}\sim\tau_{\text{d}}$.

In the DMME (Eq.(\ref{eq:nonM})), the jump operator
$\tilde F_{mn}$ denotes a transition from the state $\ket{\psi_{n}(0)}$ to another
one $\ket{\psi_{m}(0)}$. In other words, the transitions caused by the decoherence occur
between eigenstates of the LRI. Based on the Lewis-Riesenfeld phase (Eq.(\ref{eq:lrp})), the instantaneous frequency
$\alpha_{mn}^k$  can be divided into three parts, i.e.,
\begin{eqnarray}
\alpha_{mn}^k&=&-\left(\langle\psi_{m}(t)|H_{\text{s}}(t)|\psi_{m}(t)\rangle-
\langle\psi_{n}(t)|H_{\text{s}}(t)|\psi_{n}(t)\rangle\right)\nonumber\\
&&+i\left(\langle\psi_{m}(t)|\partial_{t}|\psi_{m}(t)\rangle-
\langle\psi_{n}(t)|\partial_{t}|\psi_{n}(t)\rangle\right)\nonumber\\
&&-\partial_t \text{Arg}\left(\langle\psi_{m}(t)|A_{k}|\psi_{n}(t) \rangle\right).\label{alphamn}
\end{eqnarray}
The first term in Eq.(\ref{alphamn}) attributes to a difference between the energy average values
of the eigenstates $\ket{\psi_{n}(t)}$ and $\ket{\psi_{m}(t)}$. The second term is a geometric
contribution from the time-dependent eigenstates, while the third term comes from the phase changing
rate in the transitions caused by the interaction Hamiltonian. In the adiabatic limit, the eigenstates
of the LRI are the eigenstates of the system Hamiltonian, and the adiabatic condition
must be satisfied. Thus, the last two terms are no contributions to the instantaneous frequency,
while the first term becomes the energy gap between the $n$-th and the $m$-th energy levels,
which leads to the adiabatic master equation given in Ref. \cite{Albash2012,Kamleitner2013}.

\section{The Driven Open Two-Level System} \label{sec:2qubits}

In this section, we apply the general formulism to a driven two-level system which couples
with a heat reservoir. Here, we consider that the driven two-level system Hamiltonian
in a laser adapted interaction picture takes the form \cite{Chen2010}
\begin{equation}
H_{\text{s}}(t)=\Delta(t)\sigma_{z}+\Omega(t)\sigma_{x},
\label{eq:Hs}
\end{equation}
where $\Delta(t)=\omega_0(t)-\omega_L$ is the time-dependent detuning with the time-dependent
Rabi frequency $\omega_0(t)$ and a constant laser frequency $\omega_L$; $\Omega(t)$ is
time-dependent driven field. The heat reservoir  can be represented by the reservoir Hamiltonian
$$H_\text{B}=\sum_k\Omega_k b_{k}^{\dagger}b_{k}$$
with $\Omega_k=\omega_k-\omega_L$, {where $b_{k}$ and $\omega_k$ are the annihilation
operator and the eigen-frequency of the $k$-th mode of the reservoir \cite{Shen2014}.}
Without loss of generality, the interaction Hamiltonian is selected as
\begin{equation}
H_{\text{I}}=\sum_{j=x,y}A^j \otimes B^j,
\label{eq:Hi}
\end{equation}
 where the system and bath operators are
\begin{eqnarray}
A^x=\sigma_{x},\:B^x=\sum_k g_k^x(b_{k}^{\dagger}+b_{k}),\nonumber\\
A^y=\sigma_{y},\:B^y=\sum_k i g_k^y(b_{k}-b_{k}^{\dagger}).\label{eq:bk}
\end{eqnarray}

For the two-level system governed by the Hamiltonian Eq.(\ref{eq:Hs}), the LRIs have been
explored before \cite{Chen2011,Lai1996}.  Here, we write the LRIs of the two-level system
in form of the spectrum decomposition
\begin{eqnarray}
I_{\text{s}}(t) & = &\sum_{k=1,2}\pm\Omega_\text{I}\ket{\psi_k(t)}\bra{\psi_k(t)},\label{eq:Is}
\end{eqnarray}
where $\pm\Omega_\text{I}$ are constant eigenvalues and
\begin{eqnarray}
\ket{\psi_{1}(t)} & = & \left(\cos\eta(t)\text{e}^{i\zeta(t)},\sin\eta(t)\right)^{\text{T}},\nonumber\\
\ket{\psi_{2}(t)} & = & \left(\sin\eta(t)\text{e}^{i\zeta(t)},-\cos\eta(t)\right)^{\text{T}},\label{eq:eigLRI}
\end{eqnarray}
are the eigenstates of the LRI (Eq.(\ref{eq:Is})), correspondingly.
 Inserting Eqs. (\ref{eq:Hs}) and (\ref{eq:Is}) into Eq.(\ref{eq:dIs}), the parameters
$\eta(t) $ and $\zeta(t)$ needs to satisfy the following differential equation
\begin{eqnarray}
&\partial_{t}\eta=\Omega\,\sin\zeta,\nonumber\\
&\sin2\eta\,\left(2\Delta+\partial_{t}\zeta\right)=2\,\Omega\,\cos2\eta\,\cos\zeta.
\label{eq:geq}
\end{eqnarray}

In what follows, we identify the system operator $\tilde A^{x(y)}(t)$ based on the Eq.(\ref{eq:Fj}).
On the one hand, $\bra{\psi_m(t)}A^x\ket{\psi_n(t)}$ can be obtained by means of Eqs.(\ref{eq:bk})
and (\ref{eq:eigLRI})
\begin{eqnarray}
A_{11}^x & = &\sin2\eta\,\cos\zeta\,\text{e}^{i\varphi_{11}^x},\nonumber\\
A_{12} ^x& = &\sqrt{1-\sin^{2}2\eta\,\cos^{2}\zeta}\,\text{e}^{i\varphi_{12}^x},\nonumber\\
A_{21} ^x& = &\sqrt{1-\sin^{2}2\eta\,\cos^{2}\zeta}\,\text{e}^{i\varphi_{21}^x},\nonumber\\
A_{22} ^x& = &\sin2\eta\,\cos\zeta\,\text{e}^{i\varphi_{22}^x}, \label{eq:ax}
\end{eqnarray}
so do $\bra{\psi_m(t)}A^y\ket{\psi_n(t)}$, i.e.,
\begin{eqnarray}
A_{11}^y & = &\sin2\eta\,\sin\zeta\,\text{e}^{i\varphi_{11}^y},\nonumber\\
A_{12} ^y& = &\sqrt{1-\sin^{2}2\eta\,\sin^{2}\zeta}\,\text{e}^{i\varphi_{12}^y},\nonumber\\
A_{21} ^y& = &\sqrt{1-\sin^{2}2\eta\,\sin^{2}\zeta}\,\text{e}^{i\varphi_{21}^y},\nonumber\\
A_{22} ^y& = &\sin2\eta\,\sin\zeta\,\text{e}^{i\varphi_{22}^y},\label{eq:ay}
\end{eqnarray}
in which the phases are $\varphi_{11}^x=0$, $\varphi_{22}^x=\pi$,  $\varphi_{11}^y=\pi$, $\varphi_{22}^y=0$,
\begin{eqnarray*}
\tan\varphi_{12} ^x& = & -\frac{\sin\zeta}{\cos2\eta\,\cos\zeta},\,
\tan\varphi_{21} ^x = \frac{\sin\zeta}{\cos2\eta\,\cos\zeta},\\
\tan\varphi_{12}^y & = & \frac{\cos\zeta}{\cos2\eta\,\sin\zeta},\,
\tan\varphi_{21} ^y=- \frac{\cos\zeta}{\cos2\eta\,\sin\zeta},
\end{eqnarray*}
correspondingly. After substituting Eq.(\ref{eq:eigLRI}) into Eq.(\ref{eq:lrp}),  we can obtain the
Lewis-Riesenfeld phases,
\begin{eqnarray*}
\alpha_{1} & = & \int_{0}^{t}\text{d}\tau
\left(-\partial_{\tau}\zeta\cos^{2}\eta-{\Delta}\cos2\eta-\Omega\cos\zeta\sin2\eta\right),\\
\alpha_{2} & = & \int_{0}^{t}\text{d}\tau
\left(-\partial_{\tau}\zeta\sin^{2}\eta+{\Delta}\cos2\eta+\Omega\cos\zeta\sin2\eta\right).\\
\end{eqnarray*}
Thus, the  propagator of the free dynamics for the driven two-level system with the system Hamiltonian
Eq.(\ref{eq:Hs}) can be written down  explicitly according to Eq.(\ref{eq:us}). From Eqs. (\ref{eq:ax}) and (\ref{eq:ay}),
we have
\begin{eqnarray}
\xi_{11}^x & = & \xi_{22}^x=\sin2\eta\,\cos\zeta,\nonumber\\
\xi_{12} ^x& = & \xi_{21}^x=\sqrt{1-\sin^{2}2\eta\,\cos^{2}\zeta},\nonumber\\
\xi_{11}^y & = & \xi_{22}^y=\sin2\eta\,\sin\zeta,\nonumber\\
\xi_{12}^y & = & \xi_{21}^y=\sqrt{1-\sin^{2}2\eta\,\sin^{2}\zeta}.\label{eq:xi12}
\end{eqnarray}
and
\begin{eqnarray*}
\theta_{12}^j & = & \alpha_{2}-\alpha_{1}+{\varphi}_{12}^j,\\
\theta_{21} ^j& = & \alpha_{1}-\alpha_{2}+{\varphi}_{21}^j,\\
\theta_{11} ^j& = & \varphi_{11}^j,\,\theta_{22}^j = \varphi_{22}^j,
\end{eqnarray*}
for $j=x,y$, which result in the instantaneous frequencies as
\begin{eqnarray}
\alpha_{12}^x &=&-\alpha_{21}^x\nonumber\\
& = &- \partial_{\tau}\zeta\,\cos2\eta-2\Delta\cos2\eta-2\Omega\cos\zeta\sin2\eta\nonumber\\
&&+\frac{\partial_t\eta\,\sin2\eta\sin2\zeta+\partial_t\zeta\,\cos2\eta}{1-{\sin^22\,\eta}\,{\cos^2\zeta}},\nonumber\\
\alpha_{12}^y &=&-\alpha_{21}^y\nonumber\\
& = &- \partial_{\tau}\zeta\,\cos2\eta-2\Delta\cos2\eta-2\Omega\cos\zeta\sin2\eta\nonumber\\
&&-\frac{\partial_{t}\eta\,\sin2\eta\,\sin2\zeta-\partial_{t}\zeta\,\cos2\eta}{1-\sin^{2}2\eta\,\sin^{2}\zeta},\nonumber\\
\alpha_{11}^{x(y)}&=&\alpha_{22}^{x(y)}=0.\label{eq:alp}
\end{eqnarray}
Therefore, the system operators  $\tilde A^{x(y)}(t)$  are determined
by taking $\xi_{mn}^j$, $\theta_{mn}^j$ and $\alpha_{mn}^j$ into Eq.(\ref{eq:Fj}).

Based on the parameters provided above, we can obtain the Lamb shifts and the decoherence rates via
Eqs. (\ref{eq:gamr}) and (\ref{eq:gami}). Firstly, according to Eq.(\ref{eq:thetas-1}), we have
\begin{eqnarray*}
\varTheta_{mn}^{x(y)}(t,t-s)=\theta_{mn}^{x(y)}(t-s)- \theta_{mn}^{x(y)}(t)-\alpha_{mn}^{x(y)}(t)s,
\end{eqnarray*}
which yields
\begin{eqnarray*}
\varTheta_{12}^{x(y)}(t,t-s)&&=\int_{t-s}^{t}\text{d}\tau\,\left(\alpha_{12}^{x(y)}(\tau)-\alpha_{12}^{x(y)}(t)\right)
\end{eqnarray*}
and $\varTheta_{11}^{x(y)}(t,t-s)=\varTheta_{22}^{x(y)}(t,t-s)=0$.
Secondly, let us take the reservoir  to be in an
equilibrium state at temperature $T_R$. The correlation functions of the heat reservoir
operators satisfy
\begin{eqnarray}
\text{Tr}_{\text{B}}\left\{b_{k'}b_{k}^{\dagger}\rho_{B}\right\} & = & \delta_{k'k}(1+N_{k}),\nonumber\\
\text{Tr}_{\text{B}}\left\{b_{k'}^{\dagger}b_{k}\rho_{B}\right\}  & = & \delta_{k'k}N_{k},\nonumber\\
\text{Tr}_{\text{B}}\left\{b_{k'}b_{k}\rho_{B}\right\}  & = & 0,\nonumber\\
\text{Tr}_{\text{B}}\left\{b_{k'}^{\dagger}b_{k}^{\dagger}\rho_{B}\right\}  & = &0,\label{eq:corr}
\end{eqnarray}
where $N_{k}=\left(\exp(\omega_{k}/T_R)-1\right)^{-1}$ denotes the Planck
distribution with the reservoir temperature $T_R$.
In continuum limit, the sum over $(\textsl{g}_{k}^{x(y)})^2$ can be replaced by an
integral
\begin{eqnarray}
\sum_{k}\left(\textsl{g}_{k}^{x(y)}\right)^2\rightarrow\int_{0}^{\infty}\text{d}\omega_{k}J^{x(y)}(\omega_{k})\label{eq:gk}
\end{eqnarray}
with the spectral density function $J^{x(y)}(\omega_{k})$. Inserting Eq.(\ref{eq:bk}) into
Eq.(\ref{eq:gammak}), it yields
\begin{eqnarray}
\bar{\Lambda}^{x(y)}(\alpha)&\equiv&\sum_{k,k'}g_k^{x(y)} g_{k'}^{x(y)}\bar{\Lambda}_{kk'}^{x(y)}(\alpha)\nonumber\\
& = &\int_{0}^{\infty}\text{d}\Omega_{k}J^{x(y)}(\Omega_{k}+\omega_L)\left(N_{k}
\int_{0}^{\infty}\text{d}s\text{e}^{i\left(\alpha+\Omega_{k}\right)s}\right.\nonumber\\
 &  & \left.+(N_{k}+1)\int_{0}^{\infty}\text{d}s\text{e}^{i\left(\alpha-\Omega_{k}\right)s}\right).\label{eq:lamb}
\end{eqnarray}
with $\alpha=\alpha_{mn}^{x(y)}-\omega_{\xi}$. On making use of the formula
\begin{eqnarray}
\int_{0}^{\infty}\text{d}s\text{e}^{-i\varepsilon s}=\pi\delta(\varepsilon)
-i\text{P}\frac{1}{\varepsilon}\label{eq:delta}
\end{eqnarray}
with the Cauchy principal value P, we finally arrive at
\[
\bar{\Lambda}^{x(y)}(\alpha)=\bar{\Lambda}^{\text{R},x(y)}(\alpha)+i\bar{\Lambda}^{\text{I},x(y)}(\alpha),
\]
where
\[
\bar{\Lambda}^{\text{R},x(y)}(\alpha)=\gamma_0(\alpha)\left(N(\alpha+\omega_L)+1\right)
\]
and
\begin{eqnarray*}
&&\bar{\Lambda}^{\text{I},x(y)}(\alpha)\\
=&&\text{P}\left[\int_{0}^{\infty}\text{d}\omega_{k}
J^{x(y)}(\omega_k)\left[\frac{N(\omega_{k})+1}
{\alpha+\omega_L-\omega_{k}}+\frac{N(\omega_{k})}{\alpha-\omega_L+\omega_{k}}\right]\right].
\end{eqnarray*}
with $\gamma_0(\alpha)=\pi J(\alpha)$. After inserting $\bar{\Xi}_{mn}^{\text{c(s)},x(y)}(t,\omega_\xi)$
and $\bar{\Lambda}^{\text{R(I)},x(y)}(\alpha)$ into Eq.(\ref{eq:gamf}) and taking
the inverse Fourier transformation respect to $\omega_\xi$, the Lamb shifts and the decoherence
rates can be obtained.

Without any restriction on the driving protocol, the dynamics of the driven two-level system is governed
by the following DMME in the interaction picture,
\begin{eqnarray}
\mathcal{\tilde L}\tilde{\rho}_{\text{s}}(t)=-i\left[\tilde H_{\text{LS}}(t),
\tilde{\rho}_{\text{s}}(t)\right]+\mathcal{D}^{\text{R}}\tilde{\rho}_{\text{s}}(t)+\mathcal{D}^{\text{D}}\tilde{\rho}_{\text{s}}(t),\label{eq:imeq}
\end{eqnarray}
 with the Lamb shifts $\tilde H_{\text{LS}}(t)=\sum_{j,mn} \Gamma^{\text{I},j}_{mn}(t)\tilde F_{mn}^{\dagger}
\tilde F_{mn}$. According to Eq.(\ref{eq:alp}), the instantaneous frequency is  degenerate for $mn=\{11,\, 22\}$, which
indicates a dephasing process on $\ket{\varphi_1}$  and $\ket{\varphi_2}$ .  Therefore,  we divide the
Lindblandian into two parts with the dissipators
\begin{eqnarray}
\mathcal{D}^{\text{R }}\tilde{\rho}_{\text{s}}=&&\sum_{mn=12,21} \sum_{j=x,y}\Gamma^{\text{R},j}_{mn,mn}(t)\nonumber\\
&&\times\left(\tilde F_{mn}\tilde{\rho}_{\text{s}}\tilde F_{mn}^{\dagger}
 -\frac{1}{2}\left\{ \tilde F_{mn}^{\dagger}\tilde F_{mn},\tilde{\rho}_{\text{s}}\right\} \right),\label{eq:diss}\\
 \mathcal{D}^{\text{D}}\tilde{\rho}_{\text{s}} =&& \sum_{mn=11,22}^{m'n'=11,22} \sum_{j=x,y}\Gamma^{\text{R},j}_{mn,m'n'}(t)
 \text{e}^{i\left(\theta_{mn}^j(t)-\theta_{m'n'}^j(t)\right)}\nonumber\\
 &&\times\left(\tilde F_{m'n'}\tilde{\rho}_{\text{s}}\tilde F_{mn}^{\dagger}
 -\frac{1}{2}\left\{ \tilde F_{m'n'}^{\dagger}\tilde F_{mn},\tilde{\rho}_{\text{s}}\right\} \right),\label{eq:deph}
\end{eqnarray}
which correspond to the energy dissipation and the dephasing processes, respectively. {Here, we have used the fact
 $\alpha_{12}(t)=-\alpha_{21}(t)$ , so that the terms with $mn\neq m'n'$
 in Eq.(\ref{eq:diss}) vanish because of the secular approximation.}
 It is noteworthy  that the dephasing rates in Eq.(\ref{eq:deph}) satisfy $\Gamma^{\text{R},x(y)}_{11,11}=
\Gamma^{\text{R},x(y)}_{22,22}=-\Gamma^{\text{R},x(y)}_{11,22}= -\Gamma^{\text{R},x(y)}_{22,11}$,
due to $\theta_{11}^x = \theta_{22} ^y= \pi$,  $\theta_{22}^x  =  \theta_{11} ^y=0$. By introducing a
Hermitian operator of the  interaction picture $\tilde \Sigma_z= \tilde F_{22} -\tilde F_{11}$, the dephasing
term in Eq.(\ref{eq:imeq}) can be rewritten as
\begin{eqnarray*}
 \mathcal{D}^{\text{D}}\tilde{\rho}_{\text{s}} = \Gamma^\text{R}_{d}(t)\left(\tilde  \Sigma_z\tilde{\rho}_{\text{s}}\tilde  \Sigma_z^{\dagger}
 -\frac{1}{2}\left\{ \tilde \Sigma_z^{\dagger}\tilde  \Sigma_z,\tilde{\rho}_{\text{s}}\right\} \right),
\end{eqnarray*}
with $ \Gamma^\text{R}_{d}(t)=\Gamma^{\text{R},x}_{11,11}+\Gamma^{\text{R},y}_{11,11}$.
We further define two operators $\tilde  \Sigma_+\equiv \tilde F_{21}$  and $\tilde  \Sigma_-\equiv \tilde F_{12}$,
which fulfills
\[
\tilde  \Sigma_+=\tilde  \Sigma_-^\dagger,\,\left[\tilde  \Sigma_z,\tilde  \Sigma_+\right]=
\frac{1}{2}\tilde  \Sigma_+,\,\left[\tilde  \Sigma_z,\tilde  \Sigma_-\right]=-\frac{1}{2}\tilde  \Sigma_-.
\]
Thus, the dissipative term as shown Eq.(\ref{eq:diss}) can be reproduced as
\begin{eqnarray*}
\mathcal{D}^{\text{R }}\tilde{\rho}_{\text{s}}& =&\Gamma^\text{R}_{+}(t)\left(\tilde \Sigma_+\tilde{\rho}_{\text{s}}\tilde  \Sigma_-
 -\frac{1}{2}\left\{ \tilde  \Sigma_-\tilde  \Sigma_+,\tilde{\rho}_{\text{s}}\right\} \right)\nonumber\\
&+& \Gamma^\text{R}_{-}(t)\left(\tilde \Sigma_-\tilde{\rho}_{\text{s}}\tilde  \Sigma_+
 -\frac{1}{2}\left\{ \tilde  \Sigma_+\tilde  \Sigma_-,\tilde{\rho}_{\text{s}}\right\} \right),
\end{eqnarray*}
with $\Gamma^\text{R}_{+}\equiv\Gamma^{\text{R},x}_{21}+\Gamma^{\text{R},y}_{21}$ and
$\Gamma^\text{R}_{-}\equiv \Gamma^{\text{R},x}_{12}+\Gamma^{\text{R},y}_{12}$.
Transforming back to the Schr\"{o}dinger picture, we finally arrive at the DMME,
\begin{eqnarray}
\partial_{t}\rho_{\text{s}} & = & \mathcal{L}(t)\rho_{\text{s}}\nonumber\\
 & = & -{i}\left[H_{\text{s}}(t)+H_{\text{LS}}(t),\rho_{\text{s}}(t)\right]\nonumber\\
& +& \Gamma^\text{R}_{+}(t)\left( \Sigma_+{\rho}_{\text{s}} (t) \Sigma_-
 -\frac{1}{2}\left\{   \Sigma_-  \Sigma_+,{\rho}_{\text{s}}(t)\right\} \right)\nonumber\\
&+& \Gamma^\text{R}_{-}(t)\left( \Sigma_-{\rho}_{\text{s}}(t)  \Sigma_+
 -\frac{1}{2}\left\{   \Sigma_+  \Sigma_-,{\rho}_{\text{s}}(t)\right\} \right)\nonumber\\
 & + &\Gamma^\text{R}_{d}(t)\left[  \Sigma_z,\left[{\rho}_{\text{s}}(t),\Sigma_z\right] \right].\label{eq:smeq}
\end{eqnarray}
with the time-dependent Lindblad operators $\Sigma_k=U_{\text{s}}(t)\tilde \Sigma_kU_{\text{s}}^{\dagger}(t)$ for
$k=+,-,z$, and the Lamb shift $ H_{\text{LS}}(t)=U_{\text{s}}(t)\tilde H_{\text{LS}}(t)U_{\text{s}}^{\dagger}(t)$ .

\subsection{The Adiabatic Limit} \label{adilim}

In the adiabatic limit, the corresponding LRIs satisfy $\left[H_{\text{s}}(t),
I_{\text{s}}(t)\right]=0$, and share the same eigenstates to the system Hamiltonian.
According to Eq.(\ref{eq:geq}), if $\partial_{t}\eta(t)=\partial_t \zeta(t)=0$, it yields
$\sin\zeta=0$ and $\tan2\eta=\Omega/\Delta$.  Thus, we can write down the
eigenstates of the system Hamiltonian (Eq.(\ref{eq:Hs})) in form of Eq.(\ref{eq:eigLRI})
with
\begin{eqnarray}
\zeta=0,\,\eta=\arccos\left(-\frac{\sqrt{2}}{2}\sqrt{\frac{\sqrt{\Delta ^2+\Omega ^2}-\Delta}
{\sqrt{\Delta ^2+\Omega ^2}}}\right).\label{eq:eta2}
\end{eqnarray}
It can be verified that
\[
H_{\text{s}}(t)\ket{\psi_{i}(t)}=\epsilon_{i}(t)\ket{\psi_{i}(t)}
\]
with the eigenvalues of the system Hamiltonian $\epsilon_{1,2}(t)=\mp
\sqrt{\Delta^2+\Omega^2}/2$. In such a case, the propagator can be
represented in terms of the instantaneous eigenstates of the system Hamiltonian as
shown in Eq.(\ref{eq:us}). The phases in the propagator come back to  a sum of the geometric
phases and the dynamical phases.

 Here, we consider the situation where $\textsl{g}_k^x=0$ in the interaction Hamiltonian
 Eq.(\ref{eq:Hi}) for all $k$. Thus, the expansion coefficients in Eq.(\ref{eq:Fj}) are
\begin{eqnarray*}
\xi_{11} ^y& = & \xi_{22}^y=|\sin2\eta\,\sin\zeta|=0,\\
\xi_{12} ^y& = & \xi_{21}^y=\sqrt{1-\sin^{2}2\eta\,\sin^{2}\zeta}=1,
\end{eqnarray*}
with the phase $\varphi_{12}^y=-\varphi_{21}^y=\pi/2$. Due to $\partial_t\zeta\rightarrow0$,
the geometric phases  in $\alpha_{1}$ and $\alpha_2$ are much smaller than the coresponding
dynamical phases, so that the phase in Eq.(\ref{eq:Fj}) reads
\begin{eqnarray*}
\theta_{12} ^y& = & \alpha_{2}-\alpha_{1}\\
 & = &-\int_{0}^{t}\text{d}\tau\sqrt{\Delta(\tau)^2+\Omega(\tau)^2}+\pi,
 \end{eqnarray*}
and $\theta_{21}^y=-\theta_{12}^y$, whose derivatives are
\begin{eqnarray*}
\alpha_{12}^y & = &\sqrt{\Delta(t)^2+\Omega(t)^2},\\
\alpha_{21} ^y& =- & \sqrt{\Delta(t)^2+\Omega(t)^2},
\end{eqnarray*}
respectively.

In the adiabatic limits, the reservoir correlation time $\tau_\text{B}$ is much smaller than the
non-adiabatic timescale of the driving protocol $\tau_\text{d}$, i.e., $\tau_\text{B}\ll\tau_\text{d}$\cite{Dann2021}.
Thus the Lamb shifts and the decoherence rates can be obtained from Eq. (\ref{eq:gadi}).
By considering Eq.(\ref{eq:lamb}), it yields
\begin{eqnarray}
\Gamma_{mn}^{\text{R},y}(t) & = & 2\gamma_0(\alpha_{mn})\left(N(\alpha_{mn})+1\right),
\end{eqnarray}
with $mn=12,\,21$. Note that No matter $\Delta(t)$ and $\Omega(t)$ are either positive or
negative, $\alpha_{12}^y$ ($\alpha_{21}^y$) is always positive (negative), and the Planck
distribution satisfies $N(-\alpha_{mn})=-\left(N(\alpha_{mn})+1\right)$. Therefore, the
adiabatic Markovian master equation (AME) for the driven open two-level system can
be written as \cite{Albash2012,Kamleitner2013}
\begin{eqnarray}
\partial_{t}\rho_{\text{s}} & = & \mathcal{L}(t)\rho_{\text{s}}\nonumber\\
 & = & -{i}\left[H_{\text{s}}(t)+H_{\text{LS}}(t),\rho_{\text{s}}(t)\right]\nonumber\\
&+& 2 \gamma_0(N+1)\left( \Sigma_-{\rho}_{\text{s}}(t)  \Sigma_+
 -\frac{1}{2}\left\{   \Sigma_+  \Sigma_-,{\rho}_{\text{s}}(t)\right\} \right)\nonumber\\
 & +&2 \gamma_0N\left( \Sigma_+{\rho}_{\text{s}} (t) \Sigma_-
 -\frac{1}{2}\left\{   \Sigma_-  \Sigma_+,{\rho}_{\text{s}}(t)\right\} \right).
\end{eqnarray}

\subsection{The Inertial Limit}

A NAME based on the inertial theorem has been proposed\cite{Dann2018,Dann2020,Dann2021},
in which the free propagator of
the closed quantum system is determined by decomposing the dynamical generator in
the Hilbert-Schmidt space into a rapidly changed scalar function and an adiabatically changed
matrix. In this subsection, we illustrate that the NAME based on the inertial theorem is
the DMME as shown in Eq.(\ref{eq:smeq}) in the inertial limits.

Besides the system Hamiltonian given by Eq.(\ref{eq:Hs}), two following additional operators
are needed to determine the free propagator, which are\cite{Dann2021}
\begin{eqnarray}
L(t)=\Omega(t)\sigma_{z}-\Delta(t)\sigma_{x},\:C(t)=\bar{\Omega}(t)\sigma_{z}\label{eq:LC}
\end{eqnarray}
with $\bar{\Omega}(t)=\sqrt{\Omega^2(t)+\Delta^2(t)}$. We may construct the Liouvillian
vector as $\vec v=\{H_{\text{s}}(t),L(t),C(t)\}$. { The inertial theorem requires that  the
adiabatic parameters for $H_\text{s}(t)$ is constant \cite{Dann2021}, i.e.,
\begin{eqnarray*}
\mu=\frac{\Omega(t)\partial_t\Delta(t)-\Delta(t)\partial_t\Omega(t)}{2\,\bar{\Omega}^3(t)}
\equiv\text{cons.}.
\end{eqnarray*}
Here, we call the dynamics which satisfies the requirement of the inertial theorem as the
dynamics  in the inertial limit. Under the inertial limit mentioned above}, we have $\partial_t
\vec w(t)=-i\bar\Omega(t)\mathcal B(\mu)\vec w(t)$ with $\vec w(t)= \frac{\bar\Omega(0)}
{\bar\Omega(t)}\vec v(t)$ and
\[\mathcal B(\mu)=i\left(
                  \begin{array}{ccc}
                    0 & \mu & 0 \\
                    -\mu & 0 & 1 \\
                    0 & -1 & 0 \\
                  \end{array}
                \right). \]
As a result, by calculating the eigenstates of $\mathcal B(\mu)$, we have the eigenoperators
of the free propagator
\begin{eqnarray}
\Sigma_x&=&\frac{1}{2\kappa^2\bar\Omega(t)}[-\mu H_\text{s}(t)-i\kappa L(t)+C(t)],\nonumber\\
\Sigma_y&=&\frac{1}{2\kappa^2\bar\Omega(t)}[-\mu H_\text{s}(t)+i\kappa L(t)+C(t)],\nonumber\\
\Sigma_z&=&\frac{1}{\kappa\bar\Omega(t)}[H_\text{s}(t)+\mu C(t)],\label{eq:Losz}
\end{eqnarray}
with $\kappa=\sqrt{1+\mu^2}$, which are the Lindblad operators in the NAME based on
the inertial theorem\cite{Dann2020}.

As discussed in Sec.\ref{sec:gf}, the Lindblad operators can be determined by the eigenstates of
the LRIs according to Eq.(\ref{eq:Fj}). For the driven open quantum systems with the system
Hamiltonian Eq.(\ref{eq:Hs}), the eigenstates of $\Sigma_z$  (Eq.(\ref{eq:Losz})) must be
the eigenstates of the LRIs defined in Eq.(\ref{eq:Is}), since $\Sigma_z$ is Hermitian. In order
to verify this correspondence, we check whether the eigenstates of $\Sigma_z$ fulfill the
differential equation for the parameters of the LRIs as shown in Eq.(\ref{eq:geq}).
Substituting Eqs.(\ref{eq:Hs}) and (\ref{eq:LC}) into Eq.(\ref{eq:Losz}), the eigenstates of
$\Sigma_z$ are obtained straightforwardly,
\begin{eqnarray}
|\varphi_{1}\rangle=\left(\begin{array}{c}
\frac{\left(i\,\mu\bar{\Omega}-\Omega\right)}{\sqrt{2\,\kappa\bar{\Omega}\left(\Delta+\kappa\bar{\Omega}\right)}}\\
\frac{\sqrt{\Delta+\kappa\bar{\Omega}}}{\sqrt{2\,\kappa\bar{\Omega}}}
\end{array}\right),\nonumber\\
|\varphi_{2}\rangle=\left(\begin{array}{c}
\frac{\left(\Omega-i\,\mu\bar{\Omega}\right)}{\,\sqrt{2\,\kappa\bar{\Omega}\left(\kappa\bar{\Omega}-\Delta\right)}}\\
\frac{\sqrt{\kappa\bar{\Omega}-\Delta}}{\sqrt{2\,\kappa\bar{\Omega}}}
\end{array}\right),\label{eq:iteig}
\end{eqnarray}
which can be parameterized as Eq.(\ref{eq:eigLRI}) with
\begin{eqnarray*}
\zeta=-\arctan\left(\frac{\mu\bar\Omega}{\Omega}\right),\,\eta=\arccos\left(-\frac{\sqrt{2}}{2}
\sqrt{\frac{\kappa\bar{\Omega}-\Delta}{\kappa\bar{\Omega}}}\right).
\end{eqnarray*}
The time derivatives of $\eta(t)$ and $\zeta(t)$ read
\begin{eqnarray*}
\partial_t\zeta&=&-\frac{2\mu^{2}\bar{\Omega}^{2}\Delta}{\mu^{2}\bar{\Omega}^{2}+\Omega^{2}}
-\frac{\Omega\bar{\Omega}}{\mu^{2}\bar{\Omega}^{2}+\Omega^{2}}\partial_{t}\mu,\\
\partial_t \eta&=&-\frac{\mu\Omega\bar{\Omega}}{\sqrt{\mu^{2}\bar{\Omega}^{2}+\Omega^{2}}}
+\frac{\mu\Delta}{2\kappa^{2}\sqrt{\mu^{2}\bar{\Omega}^{2}+\Omega^{2}}}\,\partial_{t}\mu.
\end{eqnarray*}
By taking $\zeta$, $\eta$ and their time-derivatives  into Eq.(\ref{eq:geq}), it can be verify that,
the differential equations  Eq.(\ref{eq:geq}) hold in the inertial limits, i.e., $\partial_{t}\mu/\sqrt{\mu^{2}
\bar{\Omega}^{2}+\Omega^{2}}\rightarrow 0$. In other words, $\ket{\varphi_1}$ and $\ket{\varphi_2}$
  are the eigenstates of a inertial LRI which requires $\partial_{t}\mu=0$.

In the following, we derive the inertial Markovian master equation according to the inertial LRI.
Here, we still consider that $\textsl{g}_k^x=0$ for all $k$ in the interaction Hamiltonian $H_\text{I}$.
By inserting Eq.(\ref{eq:iteig}) into Eq.(\ref{eq:Fjd}), it yields the expanding coefficients
\begin{eqnarray}
\xi_{11}  =  \xi_{22}=\frac{\mu}{\kappa},\,\xi_{12}  =  \xi_{21}=\frac{1}{\kappa},
\end{eqnarray}
and the phases
\begin{eqnarray*}
\theta_{12}	&=&	-\theta_{21}\nonumber\\
	&=&-	\int_{0}^{t}\text{d}\tau\,\frac{2\kappa\bar{\Omega}(\tau)\Omega^{2}(\tau)}
{\mu^{2}\bar{\Omega}^{2}(\tau)+\Omega^{2}(\tau)}+\varphi_{12}(t),
\end{eqnarray*}
with $\varphi_{12}=\arctan(\kappa\Omega/\mu\Delta)$. After some simple algebra, we obtain the
instantaneous frequency with a concise representation
\begin{eqnarray*}
\alpha_{12}	=-\alpha_{21}=2\kappa\bar{\Omega}(t).
\end{eqnarray*}
If we assume that the non-adiabatic timescale $\tau_\text{d}$ is great shorter than the
reservoir correlation time $\tau_\text{B}$, we can determine the Lamb shifts and
the decoherence rates according to Eq.(\ref{eq:gadi}). By the same procedure
in Sec.\ref{adilim}, we obtain the inertial master equation , which has precisely same
representation as shown in Ref. \cite{Dann2020}.

\subsection{Comparison to the  Exactly Solvable Models: the Dephasing Model}

Here, we further compare to another toy model, which is exactly solved
in the interaction picture. { With the same system Hamiltonian as shown
in Eq.(\ref{eq:Hs}), i.e.,
\begin{equation*}
H_{\text{s}}(t)=\Delta(t)\sigma_{z}+\Omega(t)\sigma_{x},
\end{equation*}
we consider a time-dependent interaction Hamiltonian
\begin{eqnarray*}
H_{\text{I}}=A(t) \otimes B,
\end{eqnarray*}
where the system and reservoir operators are
\begin{eqnarray}
A(t) & = & \sin2\eta\cos\zeta\sigma_{x}+\sin2\eta\sin\zeta\sigma_{y}+\cos2\eta\sigma_{z},\label{eq:sz}\\
B & = & \sum_{k}g_{k}\left(b_{k}^{\dagger}+b_{k}\right).\nonumber
\end{eqnarray}
$\eta$ and $\zeta$ are time-dependent parameters in the eigenstates
of the LRI (Eq.(\ref{eq:eigLRI})), which are governed by Eq.(\ref{eq:geq}). }
In this way, the Hamiltonian in the interaction picture can be written as
\[
\tilde{H}_{\text{I}}=\sigma_{z}\otimes\sum_{k}g_{k}\left(b_{k}^{\dagger}\text{e}^{i\Omega_{k}t}
+b_{k}\text{e}^{-i\Omega_{k}t}\right).
\]
 By using an unitary transformation
\begin{equation}
\tilde{V}=\exp\left[\frac{1}{2}\sigma_{z}\sum_{k}\left(\gamma_{k}b_{k}^{\dagger}
-\gamma_{k}^{*}b_{k}\right)\right]\label{eq:v}
\end{equation}
with $\gamma_{k}=2g_{k}\left(1-\text{e}^{i\Omega_{k}t}\right)/\Omega_{k}$,
the reservoir and two-level system decouple to each other, which leads
to an exactly solvable dynamics of the open two-level system\cite{Breuer2007,Uhrig2007,Yi2000}. The
detailed derivation can be found in Appendix \ref{decoh}.

Putting aside the exact dynamics of this toy model, we derive the
DMME for the open two-level system. Taking Eqs. (\ref{eq:eigLRI})
and (\ref{eq:sz}) into Eq.(\ref{eq:Fj}), it follows
that the amplitudes are $\xi_{11}=\xi_{22}=1$ and $\xi_{12}=\xi_{21}=0$
while the phases and instantaneous frequency read $\theta_{11}=\pi$,
$\theta_{22}=0$, and $\alpha_{11}=\alpha_{22}=0$. Thus we have $\Gamma_{12,12}=\Gamma_{21,21}=0$
and $\Gamma_{11,11}=\Gamma_{22,22}=-\Gamma_{11,22}=-\Gamma_{22,11}\equiv\Gamma_{\text{D}}$
according to Eq.(\ref{eq:gammamn}). In view of the reservoir
correlation functions Eq.(\ref{eq:corr}), $\Gamma_{\text{D}}$
may be taken the following form
\begin{eqnarray*}
\Gamma_{\text{D}} & = & \lim_{t\rightarrow\infty}\int_{0}^{\infty}d\Omega_{k}J\left(\Omega_{k}\right)\\
&&\times\int_{0}^{t}ds\left[\left(2N_{k}+1\right)\cos\Omega_{k}s-i\sin\Omega_{k}s\right].
\end{eqnarray*}
In case of the zero reservoir temperature, i.e., $N_{k}=0$, it yields
\begin{eqnarray*}
\Gamma_{\text{D}} & = & \lim_{t\rightarrow\infty}\left\{ \frac{\kappa\,\Omega_{c}^{2}t}
{\Omega_{c}^{2}t^{2}+1}+i\!\frac{\kappa\,\Omega_{c}^{3}t^{2}}{\Omega_{c}^{2}t^{2}+1}\right\} \\
 & = & i\,\kappa\Omega_{c},
\end{eqnarray*}
in which the following spectral density has been used \cite{Breuer2007}
\begin{equation}
J(\Omega_{k})=\kappa\Omega_{k}\exp\left(-\frac{\Omega_{k}}{\Omega_{c}}\right)\label{eq:specden2}
\end{equation}
with the cut-off frequency $\Omega_{c}$ and a dimensionless coupling
rate $\kappa$. As we see, under the Markovian approximation, the
real part of $\Gamma_{\text{D}}$ vanishes, which leads to a meaningless
DMME. Hence, we restrict the upper limit of the integration over $s$
to be $t$, but not $\infty$. The DMME in the interaction picture
can be written as
\[
\partial_{t}\tilde{\rho}_{\text{s}}(t)=-i\left[\tilde{H}_{\text{LS}}(t),\tilde{\rho}_{\text{s}}(t)\right]+\mathcal{D}^{\text{D}}\tilde{\rho}_{\text{s}}(t)
\]
 with the Lamb shift Hamiltonian $\tilde{H}_{\text{LS}}(t)=\Gamma_{\text{D}}^{\text{I}}(t)\tilde{\Sigma}_{z}^\dagger\tilde{\Sigma}_{z}$
and the dissipator
\[
\mathcal{D}^{\text{D}}\tilde{\rho}_{\text{s}}=\Gamma_{\text{D}}^{\text{R}}(t)\left(\tilde{\Sigma}_{z}\tilde{\rho}_{\text{s}}
\tilde{\Sigma}_{z}^{\dagger}-\frac{1}{2}\left\{ \tilde{\Sigma}_{z}^{\dagger}\tilde{\Sigma}_{z},\tilde{\rho}_{\text{s}}\right\} \right).
\]
The Lamb shift strength and the dephasing rate are
\[
\Gamma_{\text{D}}^{\text{I}}(t)=\frac{\kappa\,\Omega_{c}^{3}t^{2}}{\Omega_{c}^{2}t^{2}+1},\:
\Gamma_{\text{D}}^{\text{R}}(t)=\frac{\kappa\,\Omega_{c}^{2}t}{\Omega_{c}^{2}t^{2}+1},
\]
 while the Lindblad operator is $\tilde{\Sigma}_{z}=\tilde{F}_{22}-\tilde{F}_{11}$.

 \begin{figure}[htbp]
\centerline{\includegraphics[width=1.1\columnwidth]{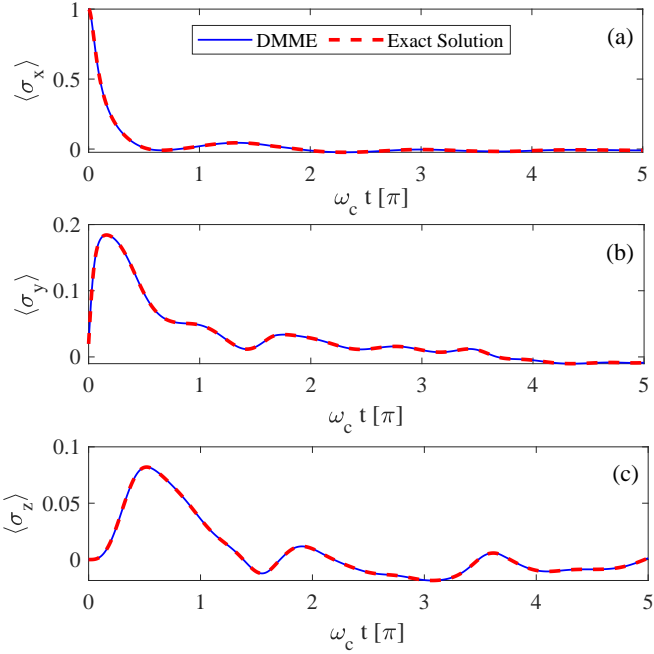}}
\caption{The main values of the Pauli operators as a function of the dimensionless time $\Omega_c t$ for
the dynamics given by the exact solution (red dashed lines), the DMME without the Lamb shift (blue solid lines),
and the DMME with the Lamb shift (green dotted lines). The parameters are chosen as $\Omega_0=\Delta=
\omega_c$, $\Omega_c=20\omega_c$, and $\kappa=1$ .  We set $\omega_c=1$ as an unity of the other
parameters.}\label{fig2}
\end{figure}

The numerical results of the main values of the Pauli operators are plotted in FIG.(\ref{fig2}).
The initial state of the system is prepared on $\rho_\text{s}(0)=(\ket{1}+\ket{0})(\bra{1}+\bra{0})/2$.
The red dashed lines are the results given by the exact solution, while the blue solid lines are
the results associated with the DMME.We consider a driving protocol with a constant detuning 
$\Delta$ and a driving field with a time-dependent strength
\begin{eqnarray*}
\Omega=\Omega_0\sin^2(\omega_c t)
\end{eqnarray*}
with constant $\Omega_0$ and $\omega_c$.
In FIG.(\ref{fig2}), we observe a strikingly good agreement between  the DMME (blue solid lines)
and the exact solution ( red dashed lines). In fact, the high uniformity does not depend on the
choice of the driving rate and the driving protocol. It is easy to verify
\[\Gamma_\text{e}(t)=\int_0^td\tau\Gamma_{\text{D}}^{\text{R}}(\tau),\]
which results in the same deocherence process for both the DMME and the exact solution.
On the other hand, due to $\tilde{\Sigma}_{z}^\dagger\tilde{\Sigma}_{z}=I$
($I$ is an identity operator), the Lamb shift does not affect the evolution
of the two level system.

\subsection{Dissipative Landau-Zener transition}

{In this subsection, we consider the dissipative Landau-Zener problem, in which the two-level
system couples to a reservoir at zero temperature. The exact transition probabilities for
such a model have been presented\cite{Wubs2006,Saito2007}. Here, we simulate the open
system dynamics of the dissipative Landau-Zener problem by the DMME, and show that the transition probability
given by the DMME almost coincides with the exact one. Meanwhile, a clear physical explanation is also
presented.}

{The dissipative Landau-Zener problem is a scattering problem in the restricted sense that changes in
the two-level system’s state will occur only during a finite time interval around $t =0$. The two-level
system's Hamiltonian has the same form as Eq.(\ref{eq:Hs}), i.e.,
\begin{equation}
H_{\text{s}}(t)=\Delta(t)\sigma_{z}+\Omega(t)\sigma_{x},\label{Eq:hlz}
\end{equation}
with $\Delta(t)=vt/2$ and  $\Omega=\Omega_0/2$, where $v$ is the constant sweep velocity and
$\Omega_0$ denotes the real intrinsic interaction amplitude between the diabatic states $\ket{1}$ and
$\ket{0}$. $\sigma_x=\ket{1}\bra{0}+\ket{0}\bra{1}$ and $\sigma_z=\ket{1}\bra{1}-\ket{0}\bra{0}$
stand for the $x$ and $z$ components of the Pauli operators. The eigenstates of the LRIs are still
given by Eq.(\ref{eq:eigLRI}), in which $\eta(t)$ and $\zeta(t)$ can be obtained via solving the differential
equations Eq.(\ref{eq:geq}) by the help of the system Hamiltonian Eq.(\ref{Eq:hlz}). Further, we assume that
 the interaction Hamiltonian takes the same form
as Eq.(\ref{eq:Hi}) with $g_k^y=0$ for all $k$, so that we obtain
\begin{eqnarray}
H_{\text{I}}=\sigma_{x}\otimes\sum_k g_k^x(b_{k}^{\dagger}+b_{k}),
\end{eqnarray}
where $b_k$ and $g_k$ are, respectively, the annihilation operator and the coupling strength of the $k$-th
mode of the reservoir.}

{With the setting above, the DMME for the  dissipative Landau-Zener problem can be written as the
same form as Eq.(\ref{eq:smeq}) with $\Gamma_{mn}^{\text{R(I)},y}=0$. For simplifying our discussion,
we neglect the Lamb shifts ($H_{\text{LS}}(t)=0$), and consider that the reservoir correlation time $\tau_{\text{B}}$
have to much smaller than the non-adiabatic phase timescale $\tau_{\text{d}}$, i.e., $\tau_{\text{B}}
\ll\tau_{\text{d}}$. According to Eq.(\ref{eq:gadi}) and by using the relations Eqs. (\ref{eq:gk}) and (\ref{eq:delta}),
the decoherence rates become
\begin{eqnarray}
\Gamma^\text{R}_{-}(t,\alpha_{12}^y)&=&\pi (\xi_{12}^y(t))^2J(\alpha_{12}^y)N(\alpha_{12}^y),\nonumber\\
\Gamma^\text{R}_{+}(t,\alpha_{21}^y)&=&\pi( \xi_{21}^y(t))^2J(\alpha_{21}^y)N(\alpha_{21}^y),
\end{eqnarray}
where $ \xi_{12}^y(t)$ and $ \xi_{21}^y(t)$ are given by Eq.(\ref{eq:xi12}). $J(\alpha)$
stands for the spectral density associated with the instantaneous frequency $\alpha_{mn}^y$,
which can be selected as
\begin{eqnarray}
 J(\alpha_{mn}^y)=\kappa\alpha_{mn}^y\exp \left(-\frac{|\alpha_{mn}^y|}{\Omega_{c}}\right)\label{eq:specden3}
 \end{eqnarray}
with the cut-off frequency $\Omega_{c}$ and a dimensionless coupling rate $\kappa$. $N(\alpha_{mn}^y)=
\left(\exp(\alpha_{mn}^y/T_R)-1\right)^{-1}$ denotes the Planck distribution with the reservoir
temperature $T_R$. Since  the Planck distribution satisfies $N(-\alpha_{mn}^y)=-\left(N(\alpha_{mn}^y)
+1\right)$ and $\alpha_{12}^y=-\alpha_{21}^y$ is always fulfilled, the instantaneous frequency
$\alpha_{12}^y$ determines the transition direction caused by decoherence. When the reservoir is
at zero temperature, it is easy to illustrate that, if $\alpha_{12}^y>0$, we have $N(\alpha_{12}^y)=1$
and $N(\alpha_{21}^y)=0$, which implies a decay from $\ket{\psi_2(t)}$ to $\ket{\psi_1(t)}$; if
$\alpha_{12}^y<0$, i.e., $\alpha_{21}^y>0$,  we have $N(\alpha_{12}^y)=0$ and $N(\alpha_{21}^y)=1$,
which leads to a decay from $\ket{\psi_1(t)}$ to $\ket{\psi_2(t)}$.
Since the instantaneous frequency $\alpha_{12}^y$ is time-dependent, the DMME may take the following
form,
\begin{eqnarray}
\partial_{t}\rho_{\text{s}}(t)
 & = & -{i}\left[H_{\text{s}}(t),\rho_{\text{s}}(t)\right]+ \mathcal D\rho_{\text{s}}(t)\label{eq:meq3}
\end{eqnarray}
with
\begin{eqnarray}
\mathcal{D}\rho_{\text{s}}=\begin{cases}
\Gamma\left(\Sigma_{-}\rho_{\text{s}}\Sigma_{+}-\frac{1}{2}\left\{ \Sigma_{+}\Sigma_{-},
\rho_{\text{s}}\right\} \right) &\text{for}\, \alpha_{12}^{y}>0\nonumber\\
\Gamma\left(\Sigma_{+}\rho_{\text{s}}\Sigma_{-}-\frac{1}{2}\left\{ \Sigma_{-}\Sigma_{+},
\rho_{+}\right\} \right) &\text{for} \,\alpha_{12}^{y}<0,\nonumber
\end{cases}
\end{eqnarray}
where $\Gamma=\kappa\pi (\xi_{12}^y(t))^2|\alpha_{12}^y|\exp \left(-{|\alpha_{12}^y|}/{\Omega_{c}}\right)$
is the decoherence rate depending on $t$ and $|\alpha_{12}^y|$.}

 \begin{figure}[htbp]
\centerline{\includegraphics[width=1.1\columnwidth]{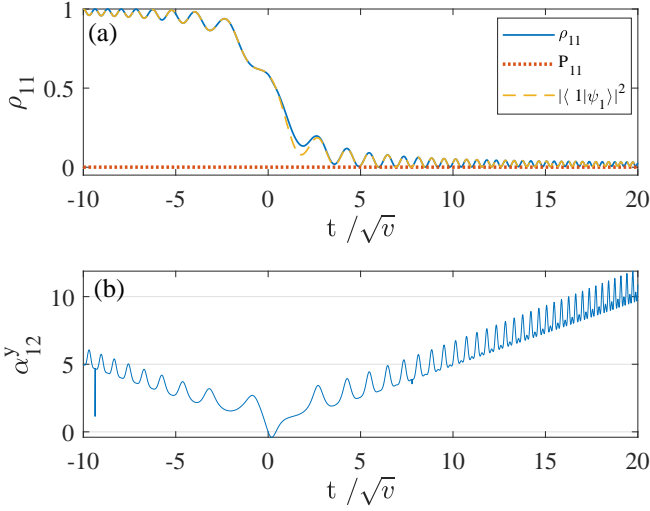}}
\caption{(a) The population on the diabatic state $\ket{1}$ given by the DMME (the blue solid line) and the Schr\"{o}dinger
Equation (the yellow dashed line), the  exact transition probability (the red dotted line) and (b) the
instantaneous frequency $\alpha_{12}^y$ as a  function of a dimensionless time $t/\sqrt{v}$.  Here we
choose $\kappa=0.1$ $\Omega_0=2/\sqrt{v}$ and $\Omega_c=8/\sqrt{v}$. We set $v=1$ as an
unity of the other parameters.}\label{fig3}
\end{figure}

{For such a dissipative Landau-Zener problem, the exact transition probability $P_{11}$
reads \cite{Wubs2006,Saito2007}
\begin{eqnarray}
P_{11}=\exp\left(-\frac{\pi W^2}{2v}\right)\label{eq:p11}
\end{eqnarray}
with
\begin{eqnarray}
W^2=\Omega_0^2+\sum_k \left(\frac{g_k^x}{2}\right)^2.
\end{eqnarray}
$P_{11}$ denotes the probability for a transition to the diabatic state $\ket{1}$, if the initial state
of the two level system is prepared in $\ket{\psi_1(-\infty)}=\ket{1}$. More directly speaking,
$P_{11}=\bra{1}\rho_\text{s}(\infty)\ket{1}$ is the population on the diabatic state $\ket{1}$ at $t=\infty$.
Considering the relation Eq.(\ref{eq:gk}), i.e., $$\sum_k (g_k^x)^2\rightarrow \int_0^\infty d\omega_k J(\omega_k)$$
with the spectral density given by Eq.(\ref{eq:specden3}), we have
\begin{eqnarray}
W^2=\Omega_0^2+\frac{\kappa}{4}\Omega_c^2.
\end{eqnarray}
In case of the closed system, the exact transition probability can be evaluated by means of
$|{\langle1|\psi_1(\infty)\rangle}|^2$, since $\ket{\psi_1(t)}$ denotes the quantum state
evolution for the closed system dynamics with the initial state $\ket{1}$.}

{In FIG. \ref{fig3} (a), we plot the evolution of the population on the diabatic state $\ket{1}$ for both the
dissipative case $\rho_{11}\equiv\bra{1}\rho_\text{s}(t)\ket{1}$ (the blue solid line) and the closed case
$|\langle1|\psi_1(t)\rangle|^2$ (the yellow dashed line). Since we have
selected $\Omega_0^2/v=4$, the adiabatic condition for the Landau-Zener transition is close
to be satisfied. Therefore, both the dissipative case and the closed case give an almost similar
dynamical evolution. On the one hand, the instantaneous frequency $\alpha_{12}^y$ is always positive
at the most of time as shown FIG. \ref{fig3} (b). Hence, the instantaneous eigenstate $\ket{\psi_1(t)}$
must be the instantaneous steady state of the DMME (Eq.(\ref{eq:meq3})). Therefore, in the adiabatic
limits, the quantum state $\rho_\text{s}(t)$ must follow with $\ket{\psi_1(t)}$. Also we find that
$\alpha_{12}^y<0$ at a small interval around $t=0$. At this time, the instantaneous steady
state becomes  $\ket{\psi_2(t)}$, so that the dissipative dynamical evolution divides from the closed
case. On the other hand, it can be observed that the  transition probability given by the DMME is very close
to the exact one $P_{11}$ (Eq.(\ref{eq:p11})). This can be understood as follows: In the adiabatic limits, i.e.,
$\Omega_0^2/v\gg 1$,  we have $\Omega_0^2\gg \sum_k (g_k^x)^2$, when the Born approximation
(the weak coupling approximation) is satisfied. Thus, the influence of  the dissipation on
the transition probability $P_{11}$ is inapparent.}

 \begin{figure}[htbp]
\centerline{\includegraphics[width=1.1\columnwidth]{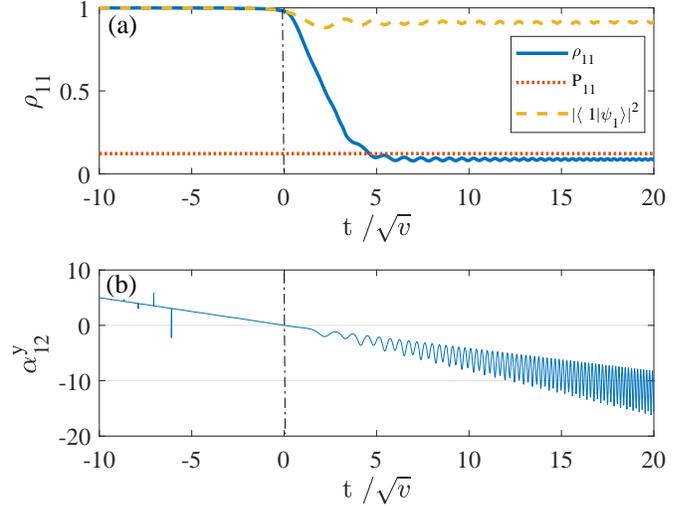}}
\caption{(a) The population on the diabatic state $\ket{1}$ given by the DMME (the blue solid line) and the Schr\"{o}dinger
Equation (the yellow dashed line), the  exact transition probability (the red dotted line) and (b) the
instantaneous frequency $\alpha_{12}^y$ as a  function of a dimensionless time $t/\sqrt{v}$.  Here we
choose $\kappa=0.1$ $\Omega_0=0.2/\sqrt{v}$ and $\Omega_c=8/\sqrt{v}$. We set $v=1$ as an
unity of the other parameters.}\label{fig4}
\end{figure}

 {The non-adiabatic case is presented in FIG. \ref{fig4}, in which we have selected $\Omega_0^2/v=0.04$.
 For the closed system case, since the adiabatic condition can not be fulfilled, the quantum state can not
 follow the eigenstates of  the Hamiltonian into the diabatic state $\ket{0}$, which are illustrated by the
 yellow dashed line in FIG. \ref{fig4} (a). When the coupling to the reservoir is considered, the result is
 entirely different. We can observe that the numerical result given by the DMME is
 in good agreement with the exact transition probability (the red dotted line). For the exact transition probability
 (Eq.(\ref{eq:p11})), due to $\Omega^2/v\ll 1$, the effect of the dissipation on the  Landau-Zener transition
 is  dominate, which reduces $P_{11}$ evidently.The decrease of $P_{11}$ can be understood by
 means of the DMME (Eq.(\ref{eq:meq3})). As shown in FIG. \ref{fig4} (b), the instantaneous frequency
 $\alpha_{12}^y(t)$ is positive at $t<0$, so that the instantaneous steady state of the DMME is the eigenstate
 of the LRI $\ket{\psi_1(t)}$. When $t>0$, the instantaneous frequency $\alpha_{12}^y(t)$  becomes
 negative. Thus, the instantaneous steady state of the DMME is turned over to the other eigenstate
 of the LRI, i.e., $\ket{\psi_2(t)}$, for $t>0$. Therefore, the population will decays into $\ket{\psi_2(t)}$
 gradually, and the final transition probability becomes $\rho_{11}(\infty)=|\langle1|\psi_2(\infty)\rangle|$.}

\section{Conclusion} \label{sec:Conc}

The driven-Markovian master equation is derived by using the LRI theory within the
Born-Markovian approximation in this paper. Since the unitary operator associated with the free
propagator of the quantum system can be decomposed by the eigenstates of the LRI, our
derivation overcomes the time-ordering obstacle in writing down a exact formula of the propagator
for the free dynamics. Due to the rapid changing of the driving protocols, the non-adiabatic
timescale may approach to, even is larger than, the reservoir correlation time, which leads
to the memory effect of the driving protocols. The DMME presented here includes this memory
effect, which leads to additional Lamb shifts and  decoherence terms. Therefore, the DMME
does not contains any constraint on the driving protocols, such as the adiabatic or inertial
approximation \cite{Albash2012,Dann2021}.

According to the DMME, the transitions of the driven open quantum system
occur on the eigenstates of the LRI, but not on the instantaneous eigenstates of the system
Hamiltonian. This is very practical in determining the Lindblad operators
in the DMME, if the LRIs are known\cite{Dann2018}, which is illustrated by
the example of the driven two-level system. Similar to the Makovian master equation with
a static Hamiltonian, both  the energy relaxation and dephasing processes emerge in the
dynamics of the driven two-level system. But the decoherence rates
and the Lindblad operators are time-dependent, which implies a time-dependent steady state.
Such a time-dependent steady state is an important candidate in the quantum state engineering
of open quantum systems. What is more, if the reservoir is at low, or ultra-low, temperature,
the steady state is  close to a pure state which is one of the eigenstates of the LRI.
Therefore, the inverse engineering method based on the LRIs\cite{Chen2010,Chen2011,Wu2021,
Tobalina2020}  will be a more promising controlling method than the others in the field
of the shortcuts to adiabaticity \cite{Campo2013,Odelin2019}.

Here, we would like to emphasise that the dynamics of the driven open quantum systems
is closely connected to the symmetry of the system, which is contained in the LRIs of
the corresponding system Hamiltonian. For instance, the DMME given in Eq.(\ref{eq:smeq}) can
describe the dynamics only for  open two-level systems with the system Hamiltonian like
Eq.(\ref{eq:Hs}). For different types of the system Hamiltonian, the open system dynamics must be
different due to distinct symmetries of the driven system. Therefore, to present the LRIs
of a particular system with a certain driving protocol is a very essential task in applying the
general formula of the DMME \cite{Ponte2018}. For driven two qubits system, a Lie-algebraic
classification and detailed construction of the LRIs has been exploited\cite{Nakahara2012}.
However, for the many body and multi-level system, this is a hard task\cite{Jezek1988}.
Fortunately, there are many useful (semi-) simple subalgebras in a complicated
Lie algebra\cite{Slansky1981}, which corresponds to many available driving
potocols. Therefore, the DMME based on the LRI theory have
broad potential applications in the quantum information process\cite{Wild2016,Albash2018}
 and the quantum thermodynamics\cite{Guarnieri2019,Vu2022}.

This work is supported by the National Natural Science Foundation of China (NSFC) under Grants
Nos. 12075050, 12175033, 12205037, and  National Key R\&D Program of China (Grant No.
2021YFE0193500). We thank Prof. J. H. An for  pointing out that the derivation of exact solution 
in the example "Comparison to the Exactly Solvable Models I:
the Dissipative Model" is wrong, which does not affect our main results.

\appendix

\begin{widetext}

\section{The Exact Evolution of the Dephasing Model }\label{decoh}

We start our derivation form taking an unitary transformation $U_{\text{s}}$
(Eq.(\ref{eq:us})) on the total Hamiltonian, which
leads to
\begin{eqnarray*}
h & = & U_{\text{s}}^{\dagger}HU_{\text{s}}\\
 & = & \sigma_{z}\otimes\sum_{k}g_{k}\left(b_{k}^{\dagger}+b_{k}\right)+\sum_{k}\Omega_{k}b_{k}^{\dagger}b_{k}.
\end{eqnarray*}
Here an interaction Hamiltonian as Eq.(\ref{eq:sz}) has been considered.
It follows that the above Hamiltonian can be exactly diagonalizated
by means of the unitary operator defined as $V=\exp\left(\sigma_{z}\otimes K\right)$
with $K=\sum_{k}\frac{g_{k}}{\Omega_{k}}\left(b_{k}^{\dagger}-b_{k}\right)$\cite{Breuer2007,Uhrig2007},
\begin{eqnarray*}
\tilde{h} & = & VhV^{\dagger}\\
 & = & \sum_{k}\Omega_{k}b_{k}^{\dagger}b_{k}-\sum_{k}\frac{g_{k}^{2}}{\Omega_{k}},
\end{eqnarray*}
in which the two-level system decouples to the heat reservoir. Mean
while, it is easy to verify that $V\sigma_{x,y}V=\sigma_{x,y}$. In
the rotating frame given by $U_{\text{s}}$, the evolution of the
system quantum state can be written down formally
\[
\tilde{\rho}(t)=U_{\text{eff}}(t)\rho(0)U_{\text{eff}}^{\dagger}(t),
\]
 where $U_{\text{eff}}(t)=\exp(-i\int_{0}^{t}h(s)ds)$ is the evolution
operator for the total system, and $\rho(t)=U_{\text{s}}\tilde{\rho}(t)U_{\text{s}}^{\dagger}$.
Thus, an useful relation for $V$ and $U_{\text{eff}}(t)$ reads
\begin{eqnarray}
U_{\text{eff}}^{\dagger}(t)V^{n}U_{\text{eff}}(t) & = & V^{\dagger}\left(VU_{\text{eff}}^{\dagger}(t)
V^{\dagger}\right)V^{n}\left(VU_{\text{eff}}(t)V^{\dagger}\right)V\nonumber \\
 & = & V^{\dagger}\exp\left(n\sigma_{z}\otimes K(t)\right)V,\label{eq:vkv}
\end{eqnarray}
with $K(t)=\exp\left(iH_{\text{B}}t\right)K\exp\left(-iH_{\text{B}}t\right)$.

Let us consider an initial product state $\rho(0)=\rho_{\text{s}}(0)\otimes\rho_{\text{B}}$
with a heat equilibrium state at temperature $T_{\text{R}}$ and a
system strate as
\[
\rho_{s}(0)=\frac{1}{2}(I+\sum_{n}r_{n}\sigma_{n}).
\]
 Here, $\sigma_{n}$ are Pauli operators and $r_{n}$ is the corresponding
component of Bloch vector satisfing $r_{n}=\text{Tr}_{\text{s}}\left\{ \rho_{s}\sigma_{n}\right\} $.
Therefore, for obtain exact evolution of the quantum state, we needs
to calculate the main values of the Pauli operators. For the x-component
in the rotating frame given by $U_{\text{s}}$, we find
\begin{eqnarray*}
\tilde{r}_{x}(t) & = & \text{Tr}\left\{ \tilde{\rho}(t)\sigma_{x}\right\} \\
 & = & \text{Tr}\left\{ \rho(0)U_{\text{eff}}^{\dagger}(t)\sigma_{x}U_{\text{eff}}(t)\right\} \\
 & = & \text{Tr}\left\{ \rho(0)U_{\text{eff}}^{\dagger}(t)V\sigma_{x}VU_{\text{eff}}(t)V^{\dagger}V\right\} ,
\end{eqnarray*}
where $V\sigma_{x}V=\sigma_{x}$ has been used. Since $VU_{\text{eff}}(t)V^{\dagger}=\exp(-i\int_{0}^{t}\tilde{h}(s)ds),$
it yields
\begin{eqnarray*}
\tilde{r}_{x}(t) & = & \text{Tr}\left\{ \rho(0)U_{\text{eff}}^{\dagger}(t)V^{2}U_{\text{eff}}(t)V^{\dagger}\sigma_{x}V\right\} \\
 & = & \text{Tr}\left\{ \rho(0)U_{\text{eff}}^{\dagger}(t)V^{2}U_{\text{eff}}(t)V^{\dagger2}\sigma_{x}\right\} .
\end{eqnarray*}
By considering Eq.(\ref{eq:vkv}), we arrive at
\begin{eqnarray*}
\tilde{r}_{x}(t) & = & \text{Tr}\left\{ \rho(0)V^{\dagger}\exp\left(2\sigma_{z}\otimes K(t)\right)V^{\dagger}\sigma_{x}\right\} \\
 & = & \text{Tr}\left\{ \left(\sigma_{x}\rho_{\text{s}}(0)\otimes\rho_{\text{B}}\right)\exp\left(2\sigma_{z}\otimes\left(K(t)-K(0)\right)\right)\right\} \\
 & = & \langle1|\sigma_{x}\rho_{\text{s}}(0)|1\rangle\left\langle \exp\left(2\left(K(t)-K(0)\right)\right)\right\rangle \\
 &  & +\langle0|\sigma_{x}\rho_{\text{s}}(0)|0\rangle\left\langle \exp\left(-2\left(K(t)-K(0)\right)\right)\right\rangle ,
\end{eqnarray*}
where $\sigma_{z}|1\rangle=|1\rangle$ and $\sigma_{x}|0\rangle=-|0\rangle$
have been used, and $\left\langle \exp\left(\pm2\left[K(t)-K(0)\right]\right)\right\rangle =\text{Tr}_{\text{B}}\left\{ \rho_{\text{B}}\exp\left(\pm2\left[K(t)-K(0)\right]\right)\right\} $
are the Wigner characteristic function of the reservoir mode $k$.
It can be easily determined by noting that it represents a Gaussian
function, which immediately leads to
\begin{eqnarray*}
 &  & \left\langle \exp\left(\pm2\left(K(t)-K(0)\right)\right)\right\rangle \\
 & = & \exp\left(2\left\langle\left( K(t)-K(0)\right)^{2}\right\rangle \right)\\
 & = & \exp\left(-4\sum_{k}\frac{g_{k}^{2}}{\Omega_{k}^{2}}\left\langle 2b_{k}^{\dagger}b_{k}+1\right\rangle\left(1-\cos\Omega_{k}t\right)\right).
\end{eqnarray*}
We now perform the continuum limit of the bath modes. Introducing the
density $f(\Omega_k)$of the modes of frequency $\Omega_k$ and defining the spectral density as\cite{Breuer2007}
\[J(\Omega_k)=4f(\Omega_k)g_k^2,\]
we can write down the decoherence function
\begin{eqnarray*}
\Gamma_{\text{e}}(t) & = & \int_{0}^{\infty}d\Omega_{k}\frac{J(\Omega_{k})}{\Omega_{k}^{2}}\left(2N_{k}+1\right)\left(1-\cos\Omega_{k}t\right)\\
 & = & \frac{1}{2}\ln\left(1+\Omega_{c}^{2}t^{2}\right),
\end{eqnarray*}
where the spectral density Eq.(\ref{eq:specden2}) has been used. Therefore, we obtain
\[
\tilde{r}_{x}(t)=r_{x}(0)\exp\left(-\Gamma_{\text{e}}(t)\right),
\]
with $N_{k}=\left\langle b_{k}^{\dagger}b_{k}\right\rangle $.
With the same procedure, we can determine the y-component of the Bloch
vector, which present a similar expression as $\tilde{r}_{x}$,
\[
\tilde{r}_{y}(t)=r_{y}(0)\exp\left(-\Gamma_{\text{e}}(t)\right).
\]
 Due to $\left[\sigma_{z},U_{\text{eff}}(t)\right]=0$, we have $\tilde{r}_{z}(t)=r_{z}(0)$.
Thus, the exact quantum state evolution of the driven two-level system
in the Schr\"{o}dinger picture can be obtained by using the unitary transformation
$U_{\text{s}}(t),$
\begin{eqnarray*}
\rho(t) & = & U_{\text{s}}(t)\tilde{\rho}(t)U_{\text{s}}^{\dagger}(t)\\
 & = & \frac{1}{2}(I+\sum_{n}\tilde{r}_{n}(t)U_{\text{s}}(t)\sigma_{n}U_{\text{s}}^{\dagger}(t)).
\end{eqnarray*}

\end{widetext}


\begin{thebibliography}{0}

\bibitem{Breuer2007} H. P. Breuer and F. Petruccione, \emph{The Theory of Open Quantum Systems}
(Oxford University Press, New York, 2007).

\bibitem{Davies1974} E. B. Davies, Commun. Math. Phys. \textbf{39}, 91 (1974).

\bibitem{Gorini1976} V. Gorini, A. Kossakowski, and E. C. G. Sudarshan, J. Math. Phys. \textbf{17}, 821 (1976).

\bibitem{Lindblad1976} G. Lindblad, Commun. Math. Phys. \textbf{48}, 119 (1976).

\bibitem{Davies1978} E. Davies and H. Spohn, J. Stat. Phys. \textbf{19}, 511 (1978).

\bibitem{Albash2012} T. Albash, S. Boixo, D. A. Lidar, and P. Zanardi, New J. Phys. \textbf{14}, 123016 (2012).

\bibitem{Childs2001} A. M. Childs, E. Farhi, and J. Preskill, Phys. Rev. A \textbf{65}, 012322 (2001).

\bibitem{Kamleitner2013} I. Kamleitner, Phys.Rev.A \textbf{87}, 042111 (2013).

\bibitem{Sarandy2004} M. S. Sarandy, L. A. Wu, and D. A. Lidar, Quant. Info. Proc. \textbf{3}, 331 (2004).

\bibitem{Yamaguchi2017} M. Yamaguchi, T. Yuge, and T. Ogawa, Phys. Rev. E \textbf{95}, 012136 (2017).

\bibitem{Dann2018} R. Dann, A. Levy, and R. Kosloff, Phys. Rev. A \textbf{98}, 052129 (2018).

\bibitem{Potts2021} P. P. Potts, A. A. Sand Kalaee, and A. Wacker, New J. Phys. \textbf{23}, 123013 (2021).

\bibitem{Albert2016}  V. V. Albert, B. Bradlyn, M. Fraas, and L. Jiang, Phys. Rev. X \textbf{6}, 041031 (2016).

\bibitem{Scopa2019} S. Scopa, G. T. Landi, A. Hammoumi, and D. Karevski, Phys. Rev. A \textbf{99}, 022105 (2019).

\bibitem{Dann2020} R. Dann, A. Tobalina, and R. Kosloff, Phys. Rev. A \textbf{101}, 052102 (2020).

\bibitem{Dann2021} R. Dann, and R. Kosloff, Phys. Rev. Research \textbf{3}, 013064 (2021).

\bibitem{Dann2019} R. Dann, A. Tobalina, and R. Kosloff, Phys. Rev. Lett. \textbf{122}, 250402 (2019).

\bibitem{Hwang2012}  B. Hwang, and H. S. Goan, Phys. Rev. A \textbf{85}, 032321 (2012).

\bibitem{Stockburger2016} J. T. Stockburger, Euro. Phys. Lett. \textbf{115}, 40010 (2016).

\bibitem{Ozaki2021} S. Ozaki and H. Nakazato, Phys. Rev. A \textbf{103}, 053713  (2021).

\bibitem{Taran2019} G. Taran, E. Bonet, and W. Wernsdorfer, Phys. Rev. B \textbf{99}, 180408(R) (2019).

\bibitem{Lewis1968} H. R. Lewis Jr., J. Math. Phys. \textbf{9}, 1976 (1968).

\bibitem{Lewis1969} H. R. Lewis Jr. and W. B. Riesenfeld, J. Math. Phys. \textbf{10}, 1458 (1969).

\bibitem{Redfield1965} A. G. Redfield, \emph{Advances in Magnetic and Optical Resonance} (Elsevier, 1965).

\bibitem{Pedrosa2009} I. A. Pedrosa and A. Rosas, Phys. Rev. Lett. \textbf{103}, 010402 (2009).

\bibitem{Sarandy2007}  M. S. Sarandy, E. I. Duzzioni, and M. H. Y. Moussa, Phys. Rev. A 76, 052112 (2007).

\bibitem{Schuch2006} D. Schuch and M. Moshinsky, Phys. Rev. A 73, 062111 (2006).

\bibitem{Chen2011} X. Chen, E. Torrontegui, and J. G. Muga, Phys. Rev. A \textbf{83}, 062116 (2011).

\bibitem{Monteoliva1994}  D. B. Monteoliva, H. J. Korsch, and J, A, Nunez, J. Phys. A \textbf{27}, 6897 (1994).

\bibitem{Nakahara2012}  U. G\"{u}ng\"{o}rd\"{u}, Y. Wan, M. A. Fasihi, and M. Nakahara, Phys. Rev. A \textbf{86}, 062312 (2012).

\bibitem{Simeonov2016}  L. S. Simeonov and N. V. Vitanov, Phys. Rev. A \textbf{93}, 012123 (2016).

\bibitem{Kim2000} S.P. Kim, A. E. Santana, F.C, Khanna, Phys. Lett. A \textbf{272}, 46 (2000).

\bibitem{Ponte2018} M. A. de Ponte, P. M. Consoli, and M. H. Y. Moussa, Phys. Rev. A \textbf{98}, 032102 (2018).

\bibitem{Petrosky1997} T. Petrosky and I. Prigogine,\emph{ The Liouville Space Extension of Quantum Mechanics},
edited by I. Prigogine and S. A. Rice (John Wiley \& Sons, New York, 1997).

\bibitem{Chen2010}  X. Chen, I. Lizuain, A. Ruschhaupt, D. Gu\'{e}ry-Odelin, and J. G. Muga, Phys.Rev.Lett. \textbf{105}, 123003 (2010).

\bibitem{Shen2014} H. Z. Shen, M. Qin, X. M. Xiu, and X. X. Yi, Phys. Rev. A \textbf{89}, 062113 (2014).

\bibitem{Lai1996}  Y. Z. Lai, J. Q. Liang, H. J. W. M\"{u}ller-Kirsten, and J.-G. Zhou, Phys. Rev. A \textbf{53}, 3691 (1996).


\bibitem{Uhrig2007} G. Uhrig, Phys. Rev. Lett. \textbf{98}, 100504 (2007).

\bibitem{Yi2000} X. X. Yi and G. C. Guo, Phys. Rev. A \textbf{62}, 062312 (2000).

\bibitem{Wubs2006} Martijn Wubs, Keiji Saito, Sigmund Kohler, Peter H\"{a}nggi, and Yosuke Kayanuma,
Phys. Rev. Lett. \textbf{97}, 200404 (2006).

\bibitem{Saito2007} Keiji Saito, Martijn Wubs, Sigmund Kohler, Yosuke Kayanuma, and Peter H\"{a}nggi,
Phys. Rev. B \textbf{75}, 214308 (2007).

\bibitem{Wu2021} S. L. Wu, W. Ma, X. L. Huang, and X. X. Yi, Phys. Rev. Applied \textbf{16}, 044028 (2021).

\bibitem{Tobalina2020} A. Tobalina, E. Torrontegui, I. Lizuain, M. Palmero, and J. G. Muga, Phys. Rev. A \textbf{102}, 063112 (2020).

\bibitem{Campo2013}  A. del Campo, Phys. Rev. Lett. \textbf{111}, 100502 (2013).

\bibitem{Odelin2019}  D. Gu\'{e}ry-Odelin, A. Ruschhaupt, A. Kiely, E. Torrontegui, S. Mart\'{\i}nez-Garaot,
and J. G. Muga, Rev. Mod. Phys. \textbf{91}, 045001 (2019).

\bibitem{Jezek1988} E. S. Hern\'{a}ndez and D. M. Jezek, Phys. Rev. A \textbf{38}, 4455 (1988).

\bibitem{Slansky1981} R. Slansky, Phys. Rep. \textbf{79}, 1 (1981).

\bibitem{Wild2016} D. S. Wild, S. Gopalakrishnan, M. Knap, N. Y. Yao, and M. D. Lukin, Phys. Rev. Lett. \textbf{117}, 150501 (2016).

\bibitem{Albash2018} T. Albash and D. A. Lidar, Rev. Mod. Phys. \textbf{90}, 015002 (2018).

\bibitem{Guarnieri2019} G. Guarnieri, G. T. Landi, S. R. Clark, and J. Goold, Phys. Rev. Research \textbf{1}, 033021 (2019).

\bibitem{Vu2022} T. VanVu, and K. Saito, Phys. Rev. Lett. \textbf{128}, 140602 (2022).




\end{thebibliography}
\end{document}